\documentclass[sigconf, nonacm=True]{acmart}
\usepackage{array} 
\usepackage{graphicx}
\usepackage{subcaption}
\usepackage{listings}
\AtBeginDocument{%
  }

\setcopyright{rightsretained}
\copyrightyear{2025}
\acmYear{2025}
\begin{document}

\title{TagLabel: RFID Based Orientation and Material Sensing for Automated Package Inspection}


\author{David Wang}
\affiliation{%
  \institution{University of Michigan}
   \city{Ann Arbor}
   \state{Michigan}
   \country{USA}}
\email{davwan@umich.edu}

\author{Jiale Zhang}
\affiliation{%
  \institution{University of Michigan}
  \city{Ann Arbor}
  \state{Michigan}
  \country{USA}}
\email{jiale@umich.edu}

\author{Pei Zhang}
\affiliation{%
  \institution{University of Michigan}
  \city{Ann Arbor}
  \state{Michigan}
  \country{USA}}
\email{peizhang@umich.edu}
\renewcommand{\shortauthors}{Wang et al.}

\begin{abstract}
Modern logistics systems face increasing difficulty in identifying counterfeit products, fraudulent returns, and hazardous items concealed within packages, yet current package screening methods remain too slow, expensive, and impractical for widespread use. This paper presents TagLabel, an RFID based system that determines both the orientation and contents of packages using low cost passive UHF tags. By analyzing how materials change RSSI and phase, the system identifies the contents of a package without opening it. Using orientation inferred from phase differences, tag occlusion, and antenna gain patterns, the system selects the tag with the greatest occlusion for accurate material sensing. We evaluate two and three tag configurations, and show that both can deliver high orientation and material sensing performance through the use of machine learning classifiers, even in realistic RF environments. When combined into a unified pipeline, TagLabel achieves more than 80 percent accuracy across all package orientations. Because it requires only standard RFID hardware and offers fast scanning times, this approach provides a practical way to enhance package inspection and improve automation in logistics operations.
\end{abstract}




\maketitle

\section{Introduction}
Counterfeit goods, fraudulent returns, and dangerous items hidden away inside packages pose a serious challenge for the e-commerce and shipping logistics industries. In the United States, retailers lose upwards of \$7.8 billion from fraudulent returns, often in the form of substituted items, with nearly 1 in 10 returns being fraudulent \cite{zhang2023understanding}. Defective or unsafe counterfeit goods account for nearly 7\% of global trade \cite{phillips2007knockoff}, often with expensive metal components substituted for cheaper plastic alternatives. At the same time, dangerous goods such as lithium ion batteries have contributed to fires onboard both passenger and cargo aircraft, with multiple serious accidents occurring as a result \cite{rizaldy2017lithium}. Several screening methods exist for packages, but have flaws which result in only a fraction of total packages being examined. X-ray machines are often used at airports or in large logistics warehouses to check inside packages \cite{chen2009x}, but the cost and minutes long scanning process of these machines prevent wide scale deployment. An alternative is manual inspection, but this process is labor intensive and can result in unintended damage to goods \cite{bendahan2017vehicle}. In addition to content inspection, knowing the orientation of packages is crucial to shipping logistics. Many sorting systems rely on scanning barcode labels, so mis-oriented packages must be detected and rotated to have their labels visible. A better approach would be an automatic system combining these processes, capable of sensing package orientation and leveraging it to determine the contents of a package without needing to open it, with scanning times short enough to be integrated into existing logistics systems.

In this paper, we propose TagLabel, an RFID based system capable of simultaneously detecting the orientation and contents of a package using either 2 or 3 smart RFID shipping labels. By detecting how various materials interact with RFID signals, the system is capable of peering inside packages without opening them. Additionally, TagLabel is able to deliver accurate results (82\% overall) in all package orientations by dynamically selecting from multiple RFID tags placed on an item. TagLabel is the first system that unites both material and orientation based RFID sensing techniques for enhanced accuracy. Compared to X-ray and manual inspection, TagLabel is able to detect the orientation and contents of a package in seconds, while using low cost standard RFID tags.

\section{Background}
RFID is a wireless technology that enables communication between a reader and small electronic tags \cite{rukundo2025survey}. These tags either draw power from the reader’s radio signal (passive) or use their own power source (active) \cite{frith2015communicating}. In this study, we focus on using UHF (ultra-high frequency) passive RFID tags, which modulate and backscatter signals from the reader in order to return data \cite{nikitin2007overview}. In addition to reading a tag’s ID and stored data, a RFID reader can also capture properties of the returned signal. These include RSSI (Received Signal Strength Indicator), which measures signal strength, and phase angle, which indicates the position of the received waveform relative to the transmitted one \cite{martinelli2015robot}. These measurements can provide insight into the tag position, tag orientation, and what the tag's signal may have passed through.

\subsection{Material Detection}
As signals from RFID tags pass through materials, the signal can be absorbed, reflected, or scattered through interactions with the molecules inside the material \cite{pradhan2017konark}. It is possible to capture these effects by recording the RSSI mean, RSSI variance, phase angle mean, and phase angle variance, which can be fed into a classifier model to determine what material the signal passed through \cite{chen2024wireless}. 
\subsubsection{Absorption}
When RF signals transit materials with a high dielectric constant, such as dense plastics or items with high water content, such as food, RF energy is absorbed, causing a slight heating effect. This results in a decrease in mean RSSI at the reader, and a small increase in mean phase shift \cite{meng2016rfid}. Absorption is often the most noticeable effect when measuring how materials interact with RF signals, especially for non-metallic objects.
\subsubsection{Reflection}
Conductive materials like metals are able to reflect RF, resulting in multipath signals that take varying times to be travel back to the reader. This leads to constructive or destructive interference at the reader, which causes a significant increase in variance for both RSSI and phase angle \cite{wang2018modeling}. Additionally, reflective materials can also alter the mean RSSI, as signals do not penetrate into materials or lose energy from absorption. 
\subsubsection{Scattering}
Scattering occurs when RF signals travel through composite materials or materials with varying internal structure, such as a wrinkled plastic wrap or cardboard boxes. Due to differences in dielectric constant between materials inside a composite, RF takes longer or shorter paths depending on which parts of the composite they interact with \cite{omar2011electromagnetic}. This moderately increases the variance of RSSI and phase angle, while slightly decreasing mean RSSI.
\subsubsection{Material Sensing With RF Interactions}
As these three types of interactions have complex effects on RSSI and phase angle, it is difficult to analytically determine what interactions occur when RF passes through materials, especially with off the shelf retail products, which may be composed of many different materials. These highly nonlinear relationships combined with noisy real world data makes material sensing an excellent candidate for deep learning techniques, which are able to capture these relationships better than other classifiers. By using RSSI mean, RSSI variance, phase angle mean, and phase angle variance as features for neural network training, we can effectively determine what material an RFID signal passes through \cite{wang2025material}. However, in order to capture variance data, sampling is required. While commercial RFID readers can read over a hundred separate tags per second, theoretically allowing for very rapid material sensing times via single RFID/phase angle samples, previous works have shown that capturing mean and variance of these features results in up to 17\% higher accuracy \cite{wang2025material}. In our study, we utilized one second sampling periods, as we determined that this was the shortest sample length where the variance and mean of RSSI and phase angle stabilized.
\begin{figure}[htbp]
  \centering
  \includegraphics[width=\linewidth]{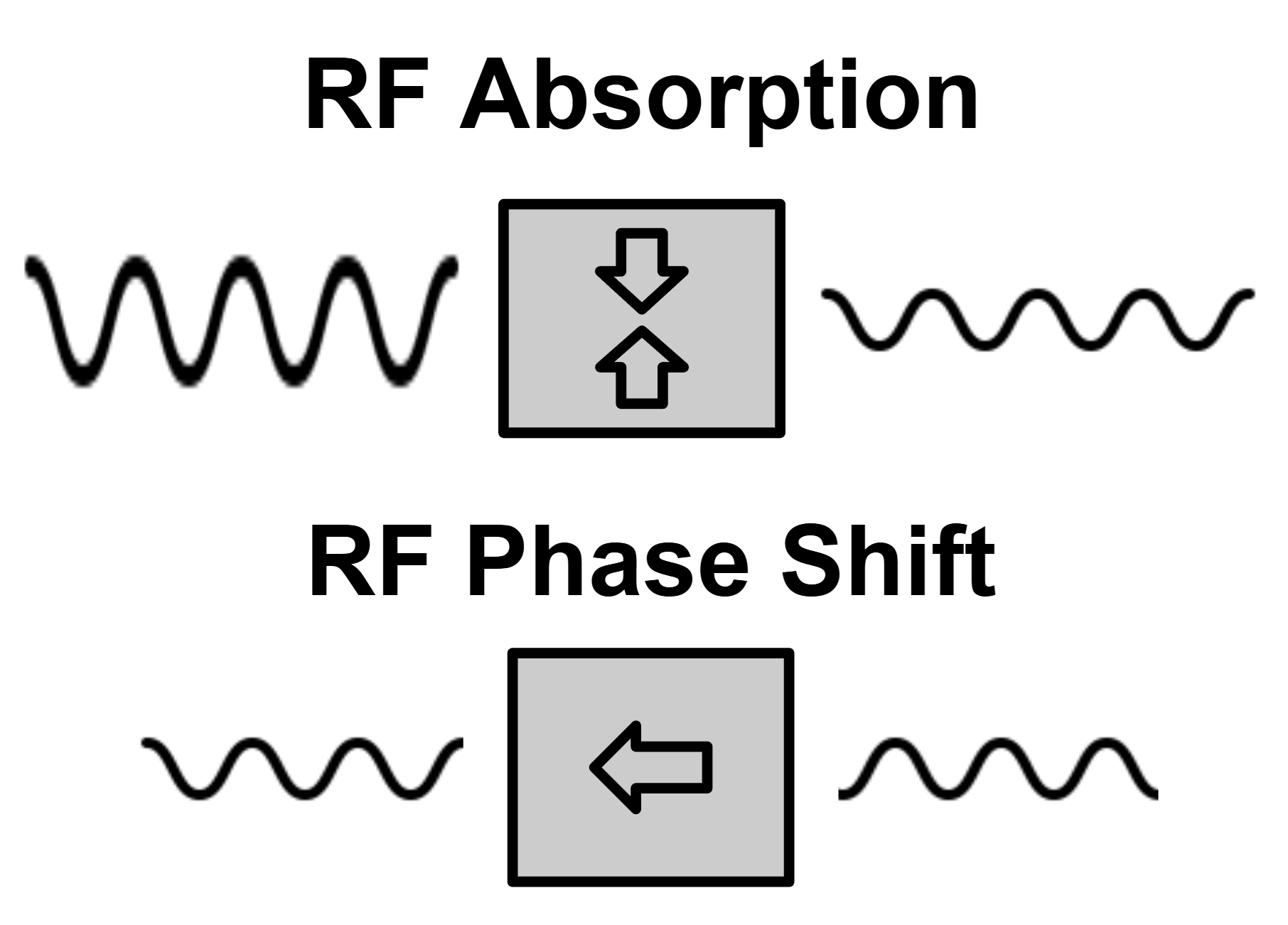}
  \caption{Interaction Phenomena Between RF and Materials}
  \Description{Interaction Phenomena Between RF and Materials}
  \label{fig:interact}
\end{figure}

\subsection{Orientation Detection}
The biggest challenge with RFID based material detection is that it relies on tags being in a specific location on an object in order for material classifiers to return an accurate result. For example, if an object has an RFID tag attached such that it is opposite the reader, RFID signals penetrate through the entire object, giving accurate material sensing results. However, if the object is then rotated such that the RFID tag now faces the reader with direct line of sight, no signal penetrates the object, making the data not useful for material sensing. This can be resolved by using multiple tags and exploiting phase differences and antenna radiation patterns to determine the orientation of an object, then picking the RFID tag with the greatest occlusion for material sensing. Knowing the orientation of packages has added benefits for logistics scenarios, as sorting systems need to rotate packages such that their barcodes are visible in order to retrieve destination information, and items need to be rotated upright throughout the shipping process to prevent damage.
\subsubsection{Phase Difference of Arrival}
When tags are placed on different sides of an object, they have a varying distance from the reader, causing different phase angles to be measured when the tag's signal returns to the reader. This can be used to calculate the Phase Difference of Arrival, or PDoA, by subtracting the phase angle of one tag from another at the receiving antenna. With the PDoAs between each tag, trigonometric relationships can be used to determine the 3D orientation of an object \cite{nikitin2010phase}. Commercially available RFID systems from manufacturers such as Impinj are able to sense phase angle at a very high level of accuracy, often within 1 degree. This makes PDoAs a very precise way of determining orientation. Analytically calculating orientation from PDoA is possible, but works best in scenarios where the object that is being measured is consistently made from the same material, since, as previously discussed, materials can cause varying shifts in phase angle. Additionally, RSSI can be used to aid PDoA based sensing, as RSSI changes with distance and occlusion, providing a second way of determining orientation \cite{peng2021deep}. A machine learning based classifier can provide more consistent results by leveraging both RSSI and phase angle as input features, and better account for noise in real world scenarios \cite{fan2019rfid}.
\subsubsection{Two Tags for Three Dimensions}
The biggest challenge with using PDoAs for sensing is that three separate tags are needed for full 3D orientation detection. In a real world deployment, this would require that 3 smart shipping labels be affixed to a package, but most packages typically have a primary address and barcode label, along with a smaller secondary barcode used for internal package routing or containing backup shipping details if the primary label is damaged. The need for a 3rd label complicates the shipping process, and 2 RFID tags can already cover all usable material sensing positions for all orientations of a rectangular package (side and rear). To sense 3D orientation with two tags, we can exploit the radiation pattern of the RFID tags. UHF RFID tags use a dipole antenna, which produces a toroidal gain pattern around the length of the tag \cite{rao2005antenna}. When the end of a tag is pointed at the reader, the RSSI decreases, while when the length of the tag is parallel to the reader, the RSSI increases, due to how the reader antenna's polarization interacts with the tag's radiation pattern. Additionally, we can make use of the fact that the bottom face of the package is against the surface it is placed on. If a tag is on this face, the RSSI is decreased drastically, as the tag is sandwiched between objects, with limited ability for multipath signals to reach the reader. By combining the PDoA between the two tags, the antenna radiation pattern, and the decreased RSSI from when the tag is on the bottom of a package, we can determine the 3D orientation using machine learning, which is able to capture the deeper relationships between these factors.
\subsubsection{Implementing Orientation Detection}
To capture PDoA and tag position related effects, we trained a random forest model, using the same one second sampling period as for material sensing. Separate models were trained for both 2 or 3 tag scenarios. The sampled RSSI and phase means were fed directly into the classifier to reduce data processing, as the PDoA between tags is a linear combination of inputs, which is easily learned when training the model.
\begin{figure}[htbp]
  \centering
  \includegraphics[width=6.5cm]{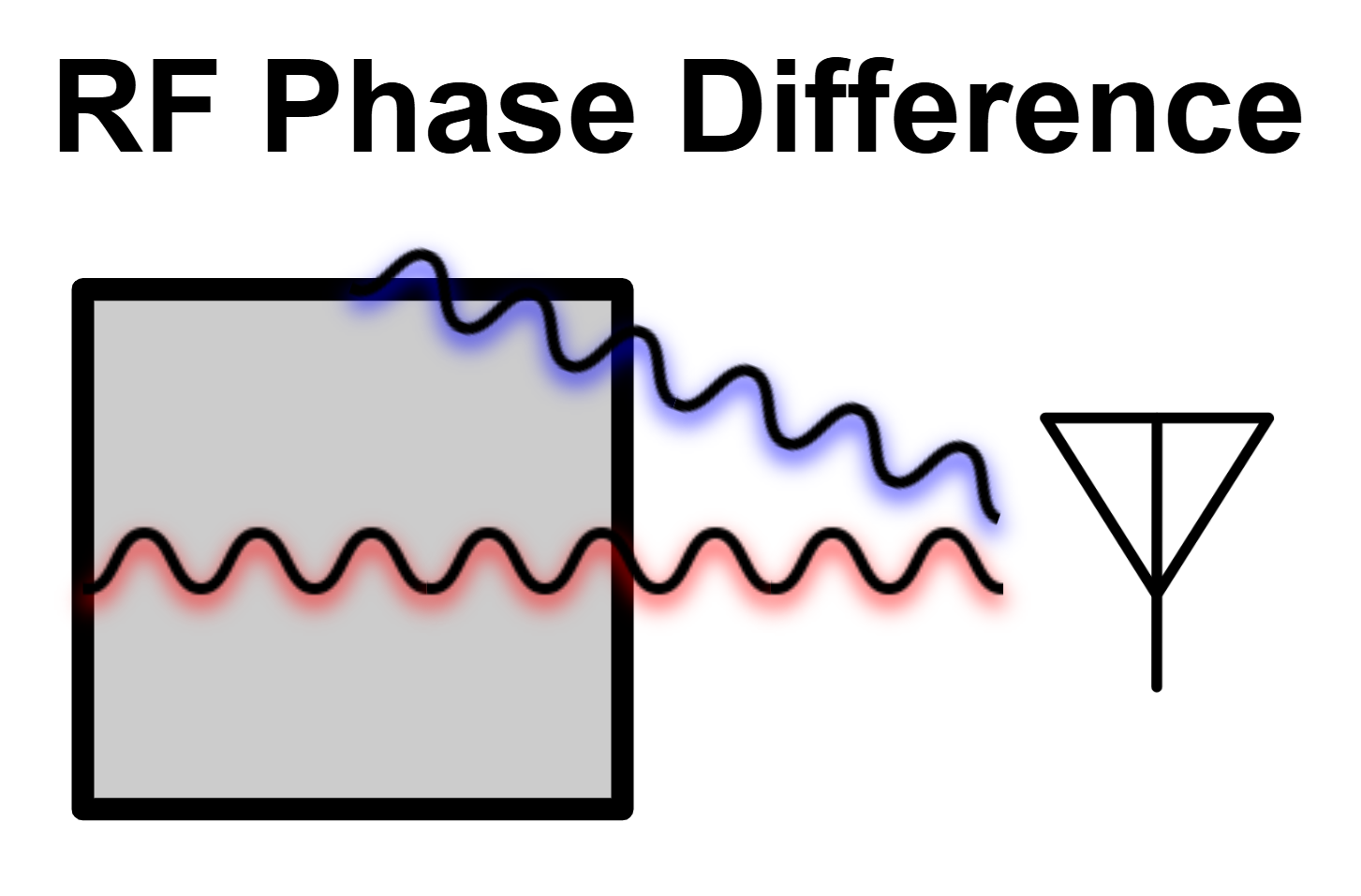}
  \caption{Illustration of Phase Difference Between Tags}
  \Description{Illustration of Phase Difference Between Tags}
  \label{fig:phase_diff}
\end{figure}
\subsection{Related Works}
Prior works have primarily focused on material and orientation as separate tasks, without the combination of both techniques to create a more robust system. In material sensing, systems such as TagTag have used tag antenna coupling with materials to detect phase changes \cite{xie2019tagtag}, but this needs the tag to be placed directly on the objects. TagScan instead use arrays of tags with objects placed in between the reader and antenna to detect RSSI and phase changes \cite{wang2017tagscan}, but this requires a much larger setup with precise placement. Our system provides a more practical mix between these two solutions with tags located on packages, simplifying the scanning process. For orientation detection, both RSSI \cite{krigslund2012orientation} and phase \cite{caccami2015phase} based approaches have been extensively explored, with some RFID sensing systems offering measurements within 2 degrees of the true orientation \cite{shirehjini2012rfid}. We leverage similar sensing techniques for our system. Even approaches that use a single RFID tag with an antenna pattern centric approach can deliver strong sensing performance \cite{liu2020pose}, so our 2 and 3 tag systems should be a robust solution for orientation sensing.

\section{System Design}
TagLabel consists of a physical RFID data collection system, and a data processing/inference pipeline with four separate classifiers. For this study, we utilized a 6x6x6 inch cardboard box as the package, and filled it with five common types of retail items: control (empty box), clothing, toilet paper, potato chips (metalized foil packaging), and plastic wrap (thick plastic film). Three UHF RFID tags were affixed to different faces of the box for use in orientation and material sensing.

\subsection{Architecture}
\begin{figure}[htbp]
  \centering
  \includegraphics[width=\linewidth]{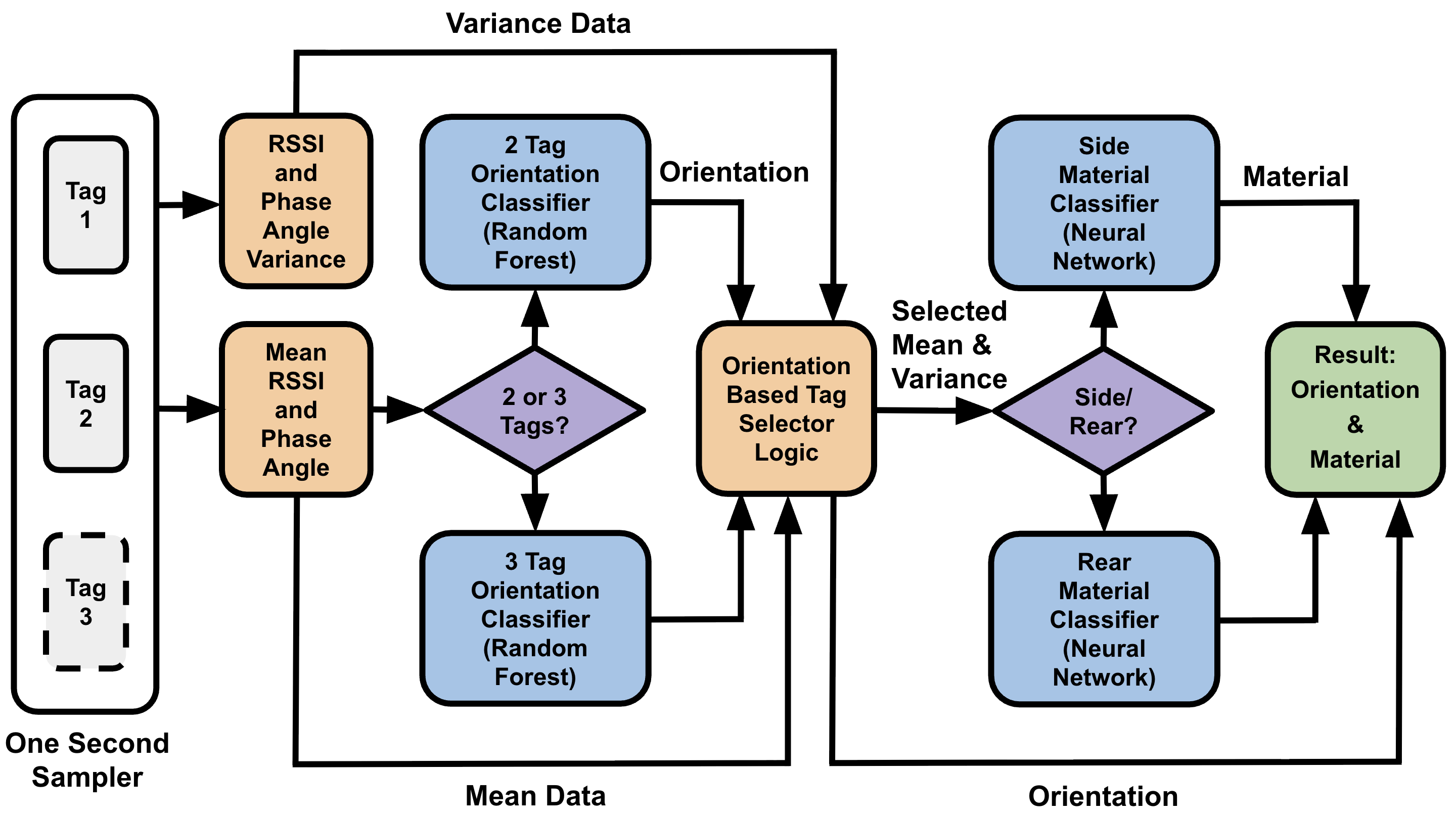}
  \caption{TagLabel Data Pipeline}
  \Description{TagLabel Data Pipeline}
  \label{fig:arch}
\end{figure}

\begin{figure}[htbp]
  \centering
  \includegraphics[width=\linewidth]{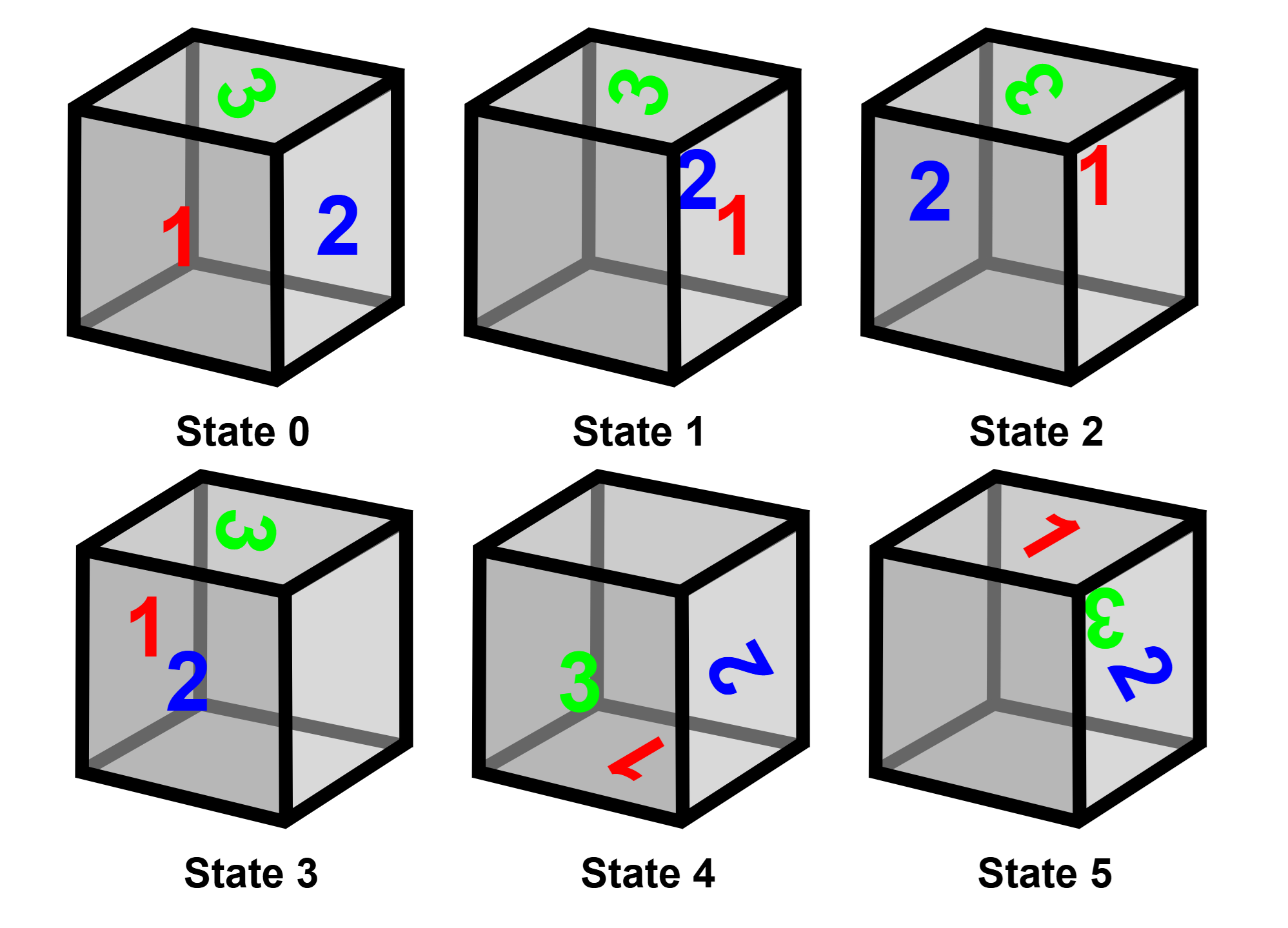}
  \caption{Tag Locations for each Orientation State}
  \Description{Tag Locations for each Orientation State}
  \label{fig:orient}
\end{figure}

\begin{figure}[htbp]
  \centering
  \includegraphics[width=\linewidth]{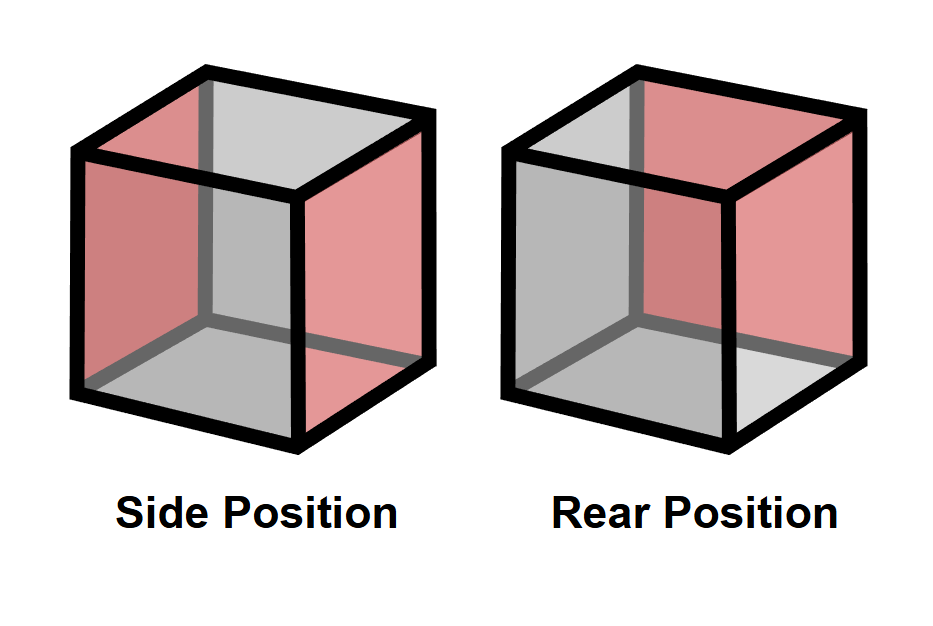}
  \caption{Tag Positions for Material Classification}
  \Description{Tag Positions for Material Classification}
  \label{fig:sides}
\end{figure}

The TagLabel system has 4 main steps, shown in Figure~\ref{fig:arch}. First, raw data collected from two or three RFID tags is processed into one second samples with mean and variance calculated for RSSI and phase angle. Depending on whether two or three tags are used, the mean RSSI and phase of the tags are fed into the respective orientation classifier, which detects one of six states, shown in Figure~\ref{fig:orient}. The orientation is then fed into the selection logic, shown in Table~\ref{tab:TagSelect}. The selection logic picks the RFID tag which has the greatest occlusion for the best material sensing performance, either a side of the package perpendicular to the reader, or the rear face of the package opposite to the reader, as seen in Figure~\ref{fig:sides}. Based on the selected tag's position on the package, the tag's mean and variance of RSSI and phase is fed into either the side or rear material classifier. Lastly, the selected classifier produces a material prediction based on the tag data.

\begin{table}[htbp]
    \centering
    \begin{tabular}{p{2.5cm}p{2.5cm}p{2.5cm}}
        \toprule
        \textbf{Orientation State} & \textbf{Selected Tag} & \textbf{Classifier} \\
        \midrule
        State 0 & Tag 2 & Side \\
        State 1 & Tag 2 & Rear \\
        State 2 & Tag 1 & Rear \\
        State 3 & Tag 1 & Side \\
        State 4 & Tag 2 & Side \\
        State 5 (3 Tags) & Tag 3 & Rear \\
        State 5 (2 Tags) & Tag 2 & Side \\
        \bottomrule
    \end{tabular}
    \caption{Tag Selection Logic}
    \Description{Tag Selection Logic}
    \label{tab:TagSelect}
\end{table}

\subsection{RFID Setup}
\begin{figure}[htbp]
  \centering
  \includegraphics[width=\linewidth]{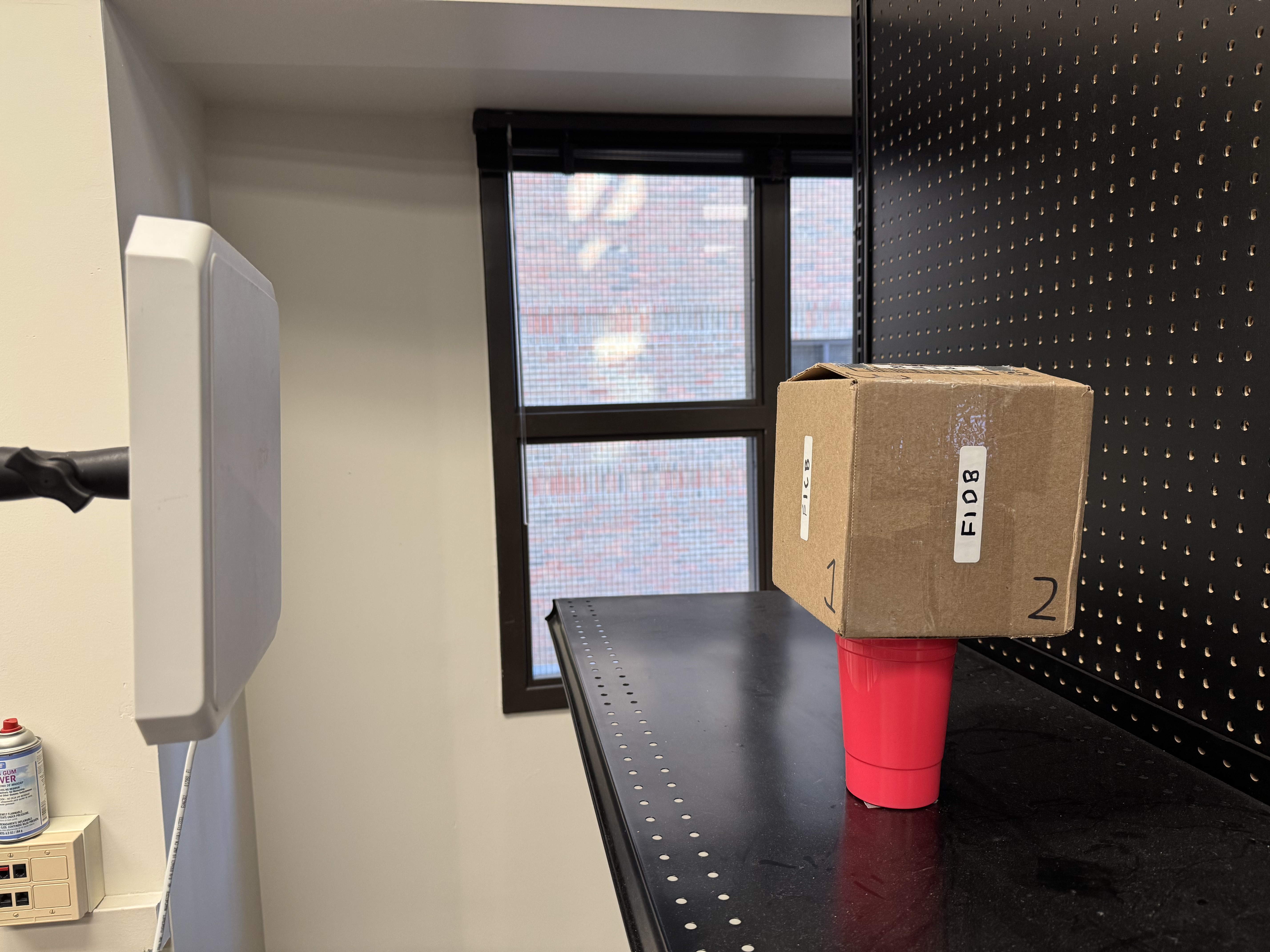}
  \caption{RFID Data Collection Setup}
  \Description{RFID Data Collection Setup}
  \label{fig:setup}
\end{figure}

The RFID setup consists of a Impinj Speedway R420 RFID reader, connected to an RFID antenna placed around 50 cm from the test package. To align the package with the center of the RFID reader, a plastic cup was placed underneath it. The cup and package were then placed on top of a metal shelf unit with a fiberboard backing. The RFID reader was configured to record RSSI and phase angle, at the maximum transmit power of 32.5 dBm, and the maximum receive sensitivity of -84 dBm. All testing was done in a mock-up cashierless retail environment to simulate realistic RF conditions, with around 55 unique RFID tagged items placed in the area surrounding the data collection point. Three UHF RFID tags with known IDs were affixed to the package, on the center of each face, with the configuration shown in Figure~\ref{fig:orient}. 
Our experimental setup is shown in Figure~\ref{fig:setup}. 

\subsection{Data Processing}
Data was collected in each of the package content and orientation scenarios for a duration of 5 minutes. This resulted in roughly 1500 individual data points for the three tags, or around 5 recorded per second. This process was repeated 6 times in all orientation states, which was then repeated for each of the 5 materials. The complete raw data contains around 45,000 individual data points. From this, the data was then processed via Pandas into 1 second samples, with mean and variance of RSSI and phase for all three tags, resulting in \textasciitilde9000 samples. These samples were split into a 8000 sample set to train/evaluate the classifiers and a 1000 sample test set to evaluate the complete TagLabel system. 

\subsubsection{Orientation Data} For the orientation classifiers, only the mean RSSI and phase from all materials was used. Two datasets with and without tag 3 were created from the classifier sample set, with inputs \lstinline|[tag1rssi,tag1phase,tag2rssi,tag2phase,tag3rssi, tag3phase]| and \lstinline|[tag1rssi,tag1phase,tag2rssi,tag2phase]| respectively, and with output \lstinline|[state]|. This resulted in \textasciitilde8000 samples for both scenarios to train and evaluate the classifiers.

\subsubsection{Material Data} For the material classifiers, two datasets were created from the classifier sample set, one for where the tag was opposite to the reader (rear) and where the tag was perpendicular to the reader (side), seen in Figure~\ref{fig:sides}. Both datasets had inputs \lstinline|[avgrssi, rssivar, avgphase, phasevar]| and output \lstinline|[material]|. As the positions of the tags change with each orientation, we used the data from each tag only when they were in the rear or side positions. Since there are two side positions compared to one rear position, the side dataset had \textasciitilde8000 samples, while the rear position had \textasciitilde4000.

\subsection{Machine Learning Classifiers}
Several machine learning techniques were tested for both the orientation and material classifiers, including logistic regression, XGBoost, SVM, random forest, and neural networks. With PCA showing that the orientation states are highly clustered and separable, we found that all types of classifiers worked well, with the ensemble decision tree based classifiers such as XGBoost and random forest performing the best. We chose to use random forests, as their accuracy was around 1-2\% better than the other options in the 2 tag scenario, while being simpler to tune. For the material classifiers, we used feedforward neural networks, as previous works \cite{fan2019rfid} have shown that they perform well in learning the relationships in RSSI and phase caused by the interactions between RF signals and various materials. To train our classifiers, we used the popular Python machine learning libraries Keras, Tensorflow, and Scikit Learn, while Pandas and NumPy were used for data manipulation. All hyperparameter tuning was done manually, with the results for both random forest and neural networks shown in our evaluation. The datasets were split into 70\% training, 15\% validation, and 15\% testing sets for the classifiers. In the random forests, we utilized bootstrap aggregation (bagging) for improved performance, and recorded the Out of Bag score (OOB) for analysis. For the neural networks, ReLU was used as the hidden layer activation function, Softmax as the output layer activation function, Adam as the optimizer, and Categorical Crossentropy as the loss function.

\subsection{Limitations}
For our experiment, we assumed that the packages would be placed a fixed distance away from the reader, as we believe this would be representative of real world deployments, such as on a conveyor belt in a processing warehouse. In all tests, we used the same 6x6x6 inch rectangular cardboard box, while in real package logistics environments, different sizes of boxes may be present, altering the orientation and material sensing responses. However, many e-commerce companies like Amazon make use of standardized box sizes, which could be linked with tag IDs to select appropriate classifiers for each size. Additionally, material and orientation accuracy performance may be higher with larger boxes, as the phase difference between tags is increased by separation, and increased interactions with materials in larger boxes can improve differentiability between material classes. Lastly, our tests only covered packages containing a single type of item, without any protective packaging apart from the cardboard box. Packages containing multiple materials may have unexpected RF interactions based on the orientation of the package, something that needs to be studied further.

\section{Evaluation}
For this experiment, we first evaluated the raw data to ensure its behavior aligns with the expected RF-material interactions and orientation detection factors. We then individually trained and assessed the performance of the orientation and material classifiers, before putting everything together to create the unified inference pipeline shown in Figure~\ref{fig:arch}.
\subsection{Raw Data: Orientations}
As discussed previously, the orientation data consists of mean RSSI and variance features either for tags 1, 2 and 3, or tags 1 and 2, depending on whether the package contains 2 or 3 tags. We used principal component analysis to project the features of both datasets onto two principal components. Each point in our PCA charts represent a single 10 second sample, colored by the orientation state as shown in Figure~\ref{fig:orient}. 
\begin{figure}[htbp]
  \centering
  \includegraphics[width=\linewidth]{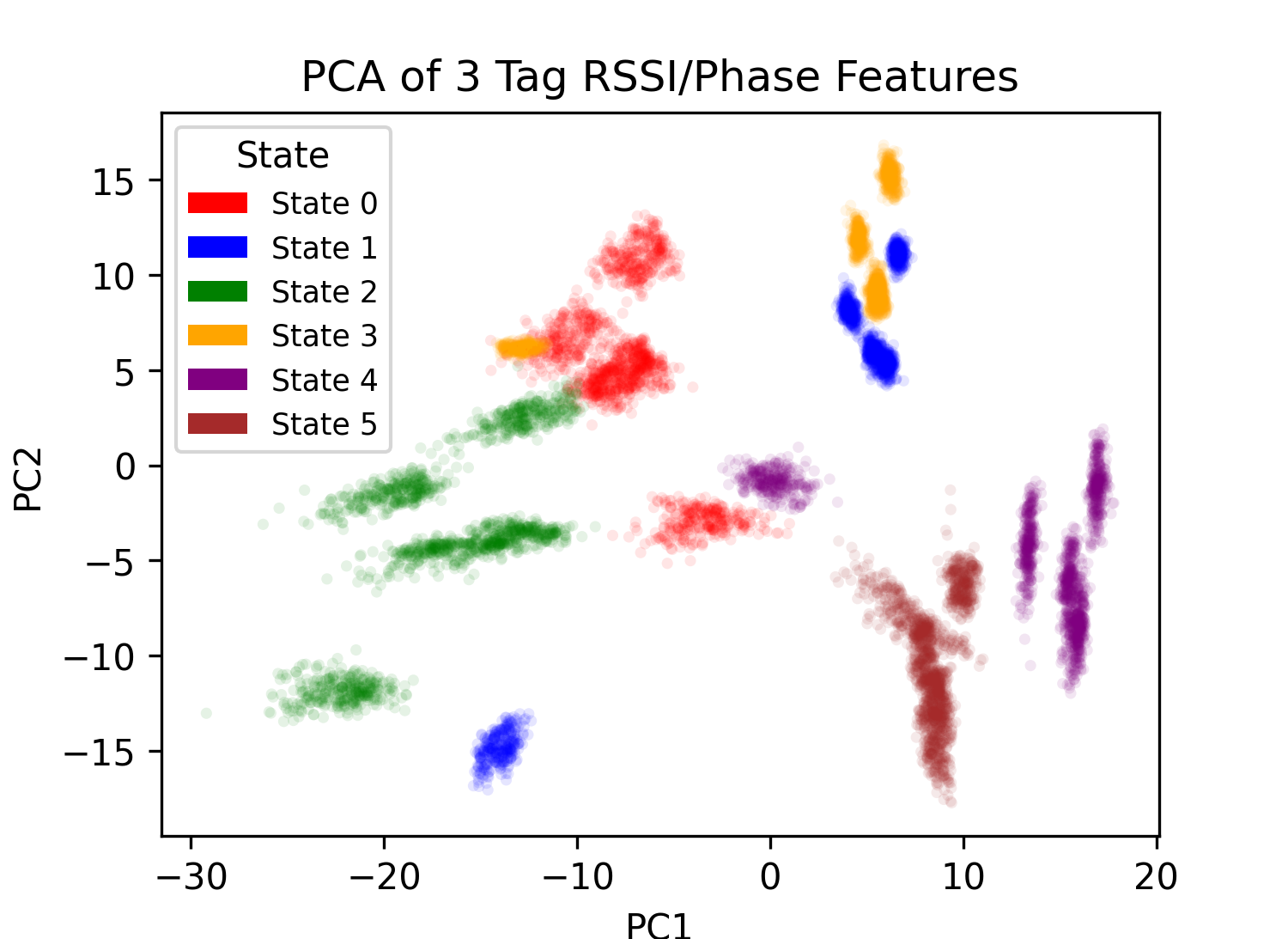}
  \caption{PCA of Features From the 3 Tag Dataset}
  \Description{PCA of Features From the 3 Tag Dataset}
  \label{fig:pca3}
\end{figure}
\begin{figure}[htbp]
  \centering
  \includegraphics[width=\linewidth]{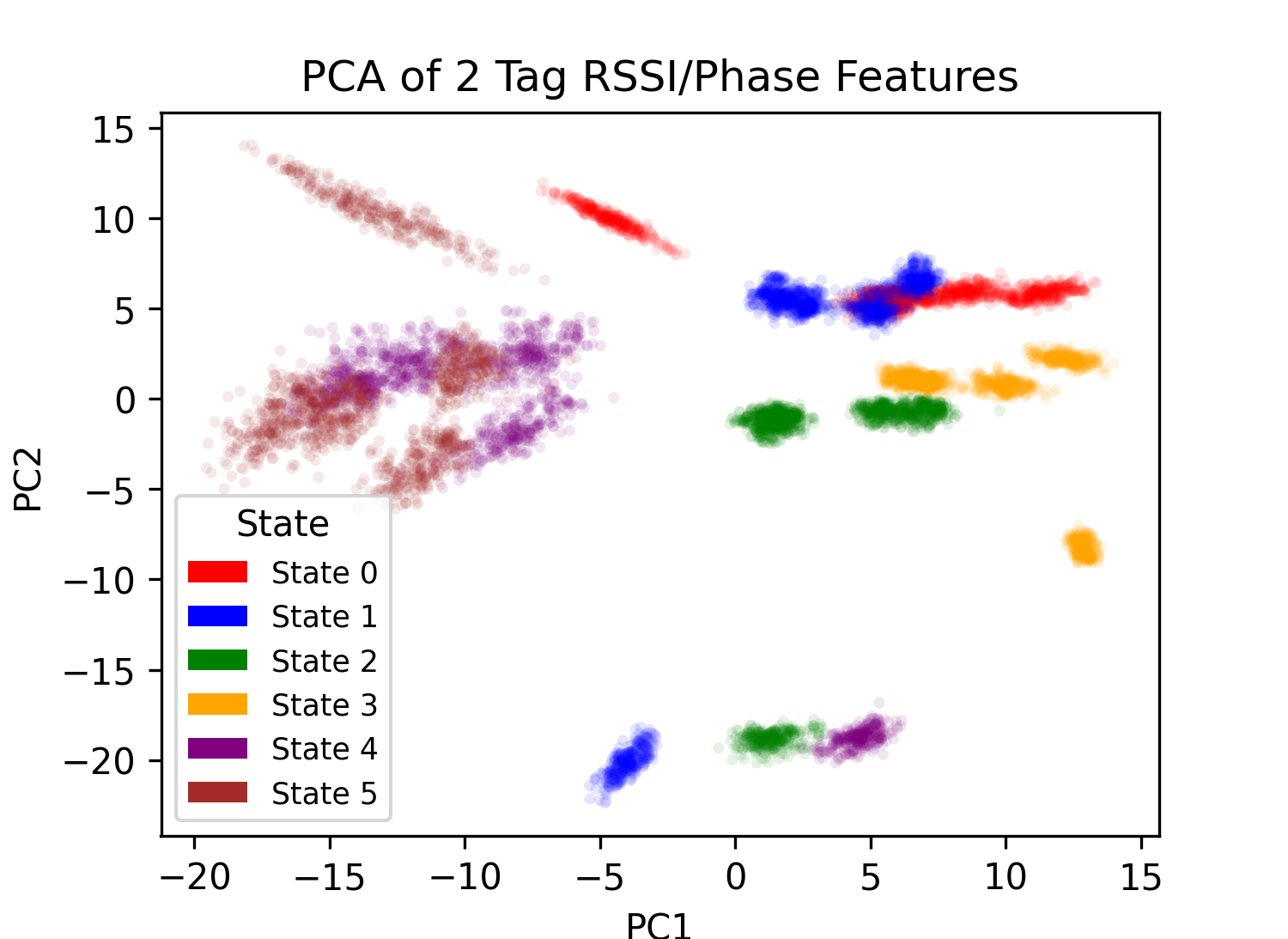}
  \caption{PCA of Features From the 2 Tag Dataset}
  \Description{PCA of Features From the 2 Tag Dataset}
  \label{fig:pca2}
\end{figure}
\subsubsection{Three Tags} The PCA for the 3 tag dataset is shown in Figure~\ref{fig:pca3}. We observed several clearly defined clusters corresponding to each state, indicating that RSSI and phase features are strongly dependent on the state. While some states appear in singular clusters, others appear in multiple clusters, showing that there may be variance among the features for that state. Due to the highly clustered and separated nature of the 3 tag dataset, we expect that any classifier for this scenario would deliver very high accuracy.
\subsubsection{Two Tags} The PCA for the 2 tag dataset is shown in Figure~\ref{fig:pca2}. While this scenario still contains several clearly defined clusters, we noted that certain cases, such as state 4 and 5, show increased variability. Additionally, we saw slight overlap between states 0 and 1, along with states 4 and 5, which may decrease classification accuracy. We also again observed that some states exhibit multiple clusters and thus may have variance in their corresponding features. We expect that any classifier for this scenario would still deliver high accuracy, but less than that of the 3 tag scenario due to the slight overlap between states.

\subsection{Raw Data: Materials}
The material data makes use of 4 features, mean and variance for RSSI and phase, which we visualized using a box plot. As the side and rear classifiers have different levels of occlusion and rely on separate classifiers, their respective data is shown separately. Each box corresponds to the full raw data for each material, totaling around 22,500 data points for the side position and 45,000 for the rear position.
\begin{figure}[htbp]
    \centering
    \begin{subfigure}[t]{7.5cm}
        \centering
        \includegraphics[width=\linewidth]{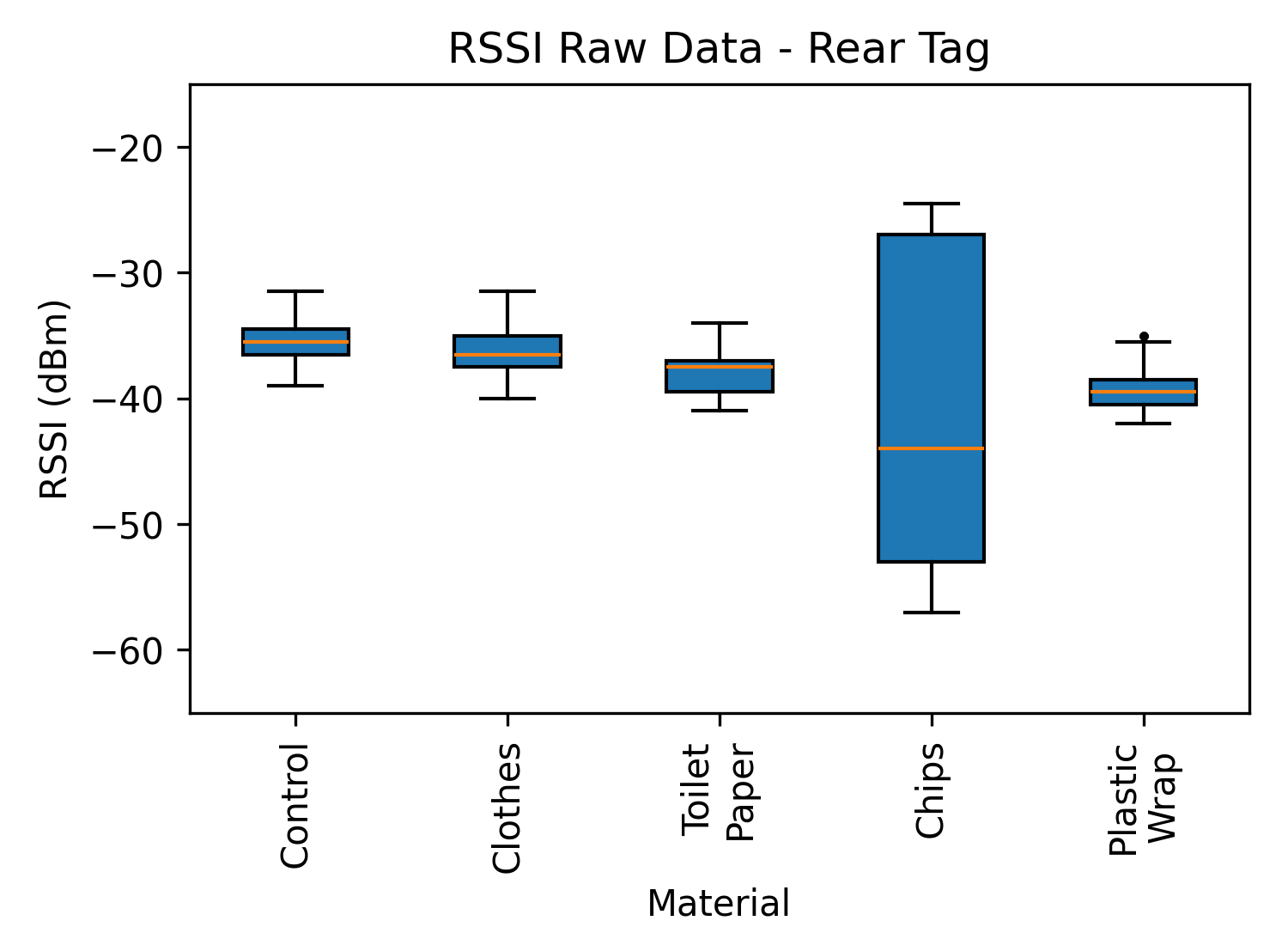}
        \caption{Observed RSSI - Rear Tag Position}
    \end{subfigure}
    \hfill
    \begin{subfigure}[t]{7.5cm}
        \centering
        \includegraphics[width=\linewidth]{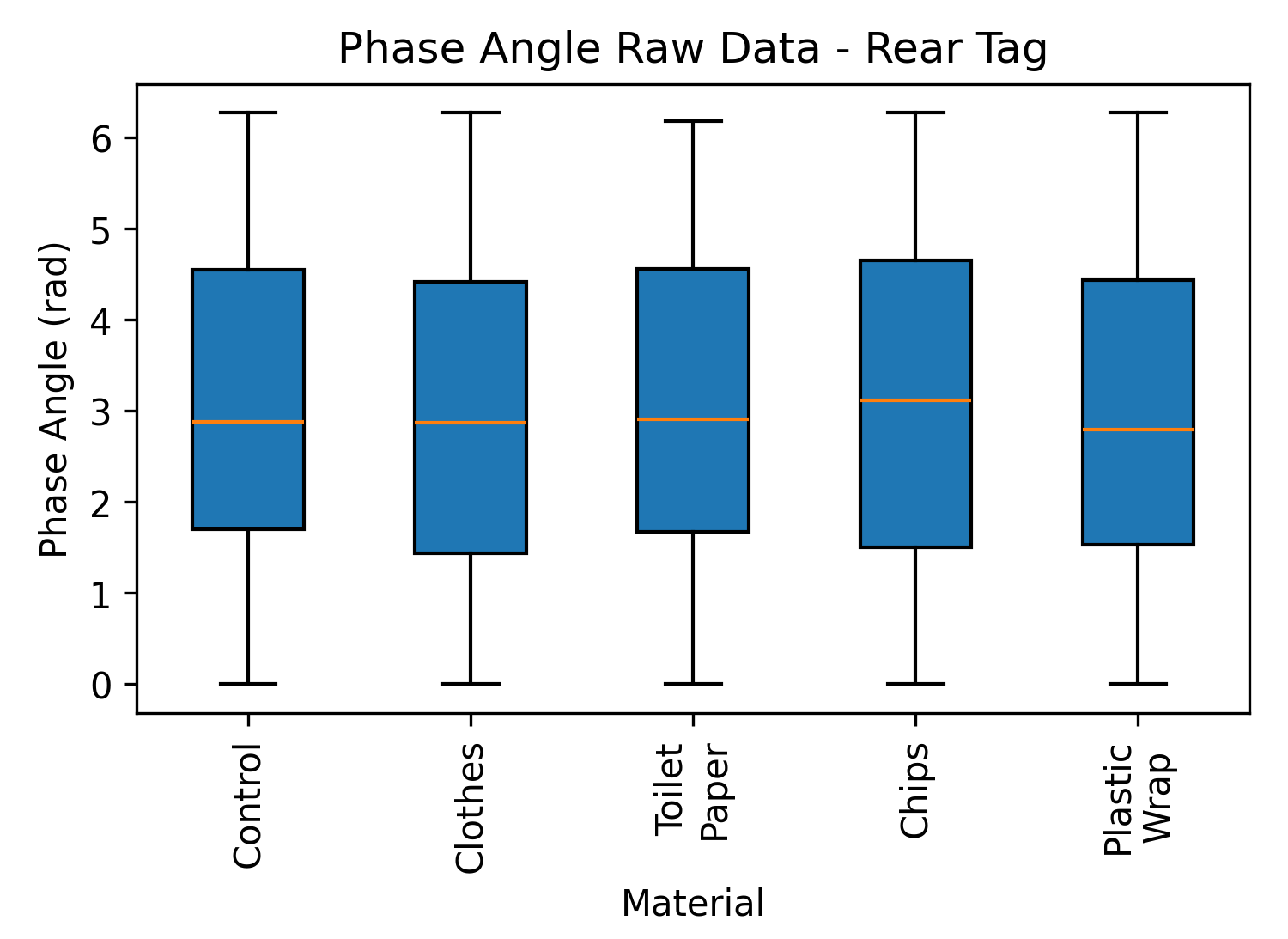}
        \caption{Observed Phase Angle - Rear Tag Position}
    \end{subfigure}
    \caption{Raw Data - Rear Tag Position}
    \label{fig:rear_raw}
\end{figure}
\begin{figure}[htbp]
    \centering
    \begin{subfigure}[t]{7.5cm}
        \centering
        \includegraphics[width=\linewidth]{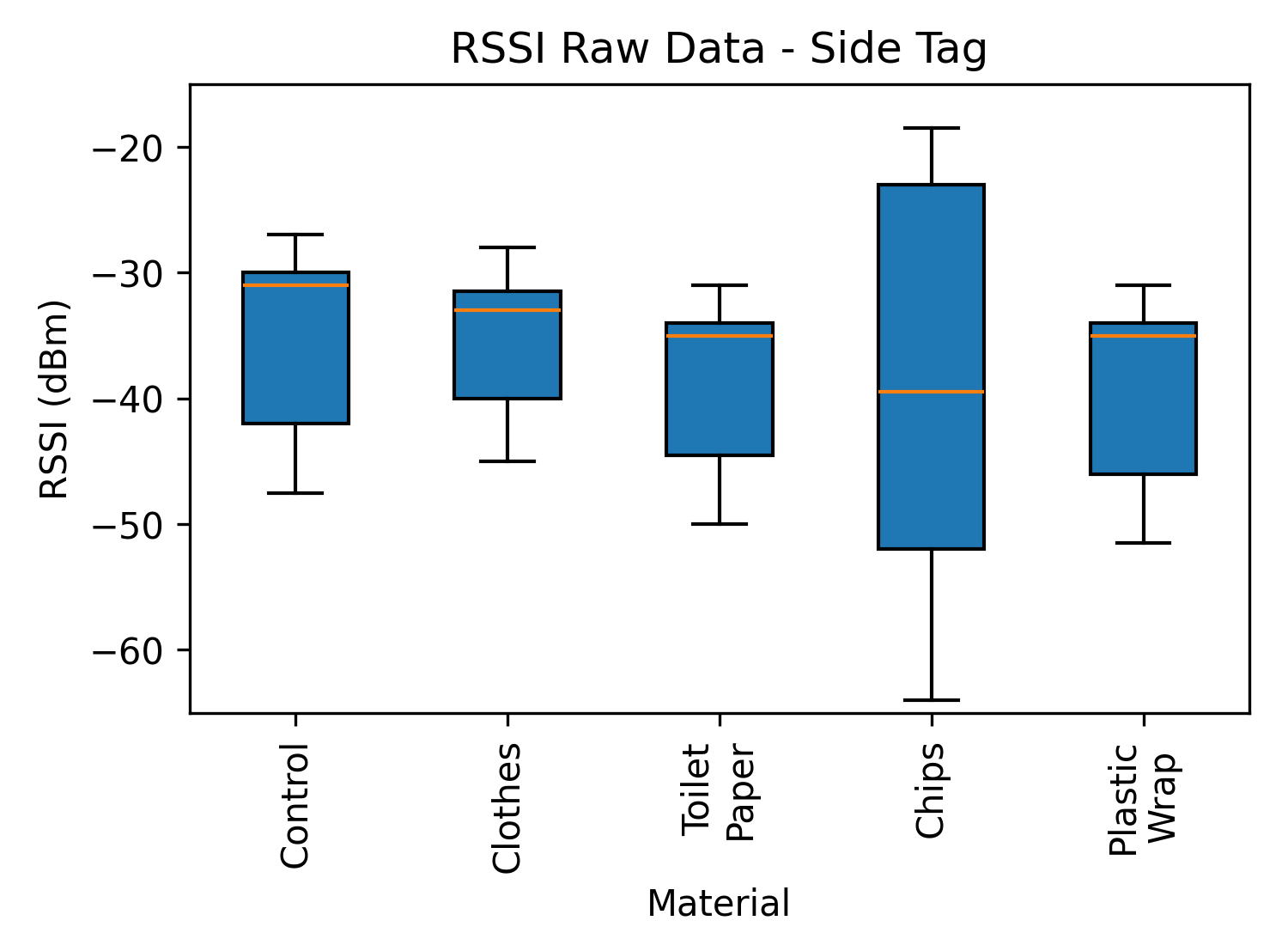}
        \caption{Observed RSSI - Side Tag Position}
    \end{subfigure}
    \hfill
    \begin{subfigure}[t]{7.5cm}
        \centering
        \includegraphics[width=\linewidth]{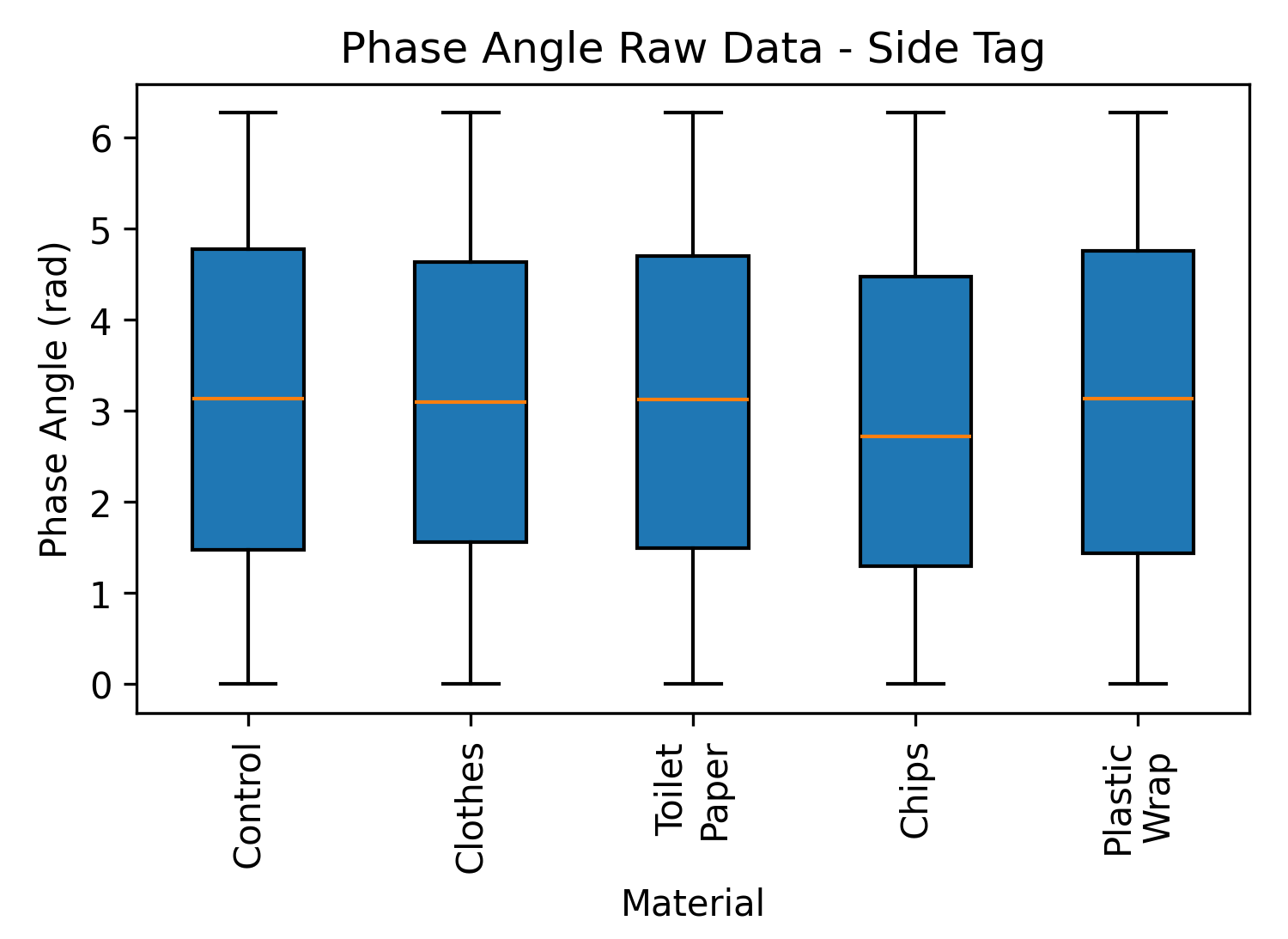}
        \caption{Observed Phase Angle - Side Tag Position}
    \end{subfigure}
    \caption{Raw Data - Side Tag Position}
    \label{fig:side_raw}
\end{figure}
\subsubsection{Rear Position}
The dataset for the rear position (Figure~\ref{fig:rear_raw}) exhibits behavior inline with RFID interaction phenomena to a strong degree. Low density materials such as the clothing exhibit a stronger RSSI, while denser materials such as the plastic wrap decrease the RSSI of the signal, consistent with RF absorption. The chips stand out from the other materials, as the non-uniform metallic packaging reflects the RFID signals wildly, resulting in multipathing that causes a huge increase in variance. When compared to the control and the more uniformly distributed toilet paper, the materials with a varying internal structure, such as the clothes and chips, exhibit a higher variance in phase, likely due to phase shifts caused by scattering. The median phase angle remains relatively constant, except for the chips, which have a slightly higher median, likely also due to their reflective metal packaging.

\subsubsection{Side Position}
The side position data (Figure~\ref{fig:side_raw}) shows some significant differences from the rear tag data. The variance in RSSI is much higher for all materials, and is skewed towards weaker RSSIs, while having a stronger median RSSI. When the tag is in the rear position, the RFID signals penetrate through the entire package before reaching the tag, but in the side position, they penetrate through less than 1/3 of the package. Additionally, due to the radiation pattern of the tag, it is much more likely that RFID signals do not pass through the package at all compared to the rear position. These factors result in the larger RSSI variance and skewed RSSI distribution observed in this data set. Compared to the rear tag position, it is harder to discern the phase angle variance between the material classes, likely also due to the decreased signal penetration. One thing to note is that the toilet paper and plastic wrap exhibit a very similar RSSI and phase distribution, which can cause confusion in classifiers. 

\subsection{Orientation Classification}
\subsubsection{Model Tuning}
Using manual hyperparameter tuning, we settled on the values shown in Table~\ref{tab:rf_arch} for our random forests, which achieved the best classification performance. As the 3 tag scenario is much easier than the 2 tag scenario, we were able to reduce the number of trees without any effect on performance. All other hyperparameters were kept the same between the two scenarios. The performance metrics achieved by both classifiers are shown in Table~\ref{tab:tv_rf}.

\begin{table}[htbp]
    \centering
    \begin{tabular}{p{4.5cm}|p{1.5cm}p{1.5cm}}
        \toprule
        \textbf{ } & \textbf{\shortstack{3 Tag\\Classifier}} & \textbf{\shortstack{2 Tag\\Classifier}} \\
        \midrule
        Features & 6 & 4 \\
        Classes & 6 & 6 \\
        Trees & 100 & 200 \\
        Max Depth & None & None \\
        Minimum Samples for Node Split & 4 & 4 \\
        Minimum Samples for Leaf Node & 2 & 2 \\
        \bottomrule
    \end{tabular}
    \caption{Random Forest Architecture}
    \Description{Random Forest Architecture}
    \label{tab:rf_arch}
\end{table}

\begin{table}[htbp]
    \centering
    \begin{tabular}{p{3.5cm}|p{2cm}p{2cm}}
        \toprule
        \textbf{ } & \textbf{\shortstack{3 Tag\\Classifier}} & \textbf{\shortstack{2 Tag\\Classifier}} \\
        \midrule
        Training Accuracy & 0.9998 & 0.9831 \\
        Validation Accuracy & 0.9992 & 0.9677 \\
        Test Accuracy & 1.0000 & 0.9713 \\
        Out of Bag (OOB) Score & 0.9995 & 0.9620 \\
        \bottomrule
    \end{tabular}
    \caption{Orientation Classifier Performance}
    \Description{Orientation Classifier Performance}
    \label{tab:tv_rf}
\end{table}

\subsubsection{Three Tag Orientation Classification}
The 3 tag scenario's random forest achieves perfect accuracy across all classes (Figure~\ref{fig:cf_rear}), with a test accuracy of 100\%. While overfitting is a concern, the PCA shows that the data is highly clustered and separate for all 6 orientation states, making it easy to differentiate, and the OOB score of 0.9995 indicates that even with unseen data, the classifier is able to identify the orientation accurately. Additionally, to ensure that data leakage caused by time dependence is not an issue, the data within each class was randomly shuffled, but this did not result in any changes to accuracy. We believe that in this specific scenario, the classification problem is simple enough to explain the 100\% accuracy, as with 3 tags, the orientation can be analytically determined with PDoAs. However, even if the model performs perfectly in this test, it is likely that in a real world scenario, the 3 tag orientation accuracy would be very high but not perfect. Given that this experiment took place in controlled environment and that the test package was not moved during each data collection period, we would expect to see less consistent results if packages were placed slightly off axis, or if RFID tags were affixed differently across packages.

\subsubsection{Two Tag Orientation Classification}
\begin{figure}[htbp]
  \centering
  \includegraphics[width=\linewidth]{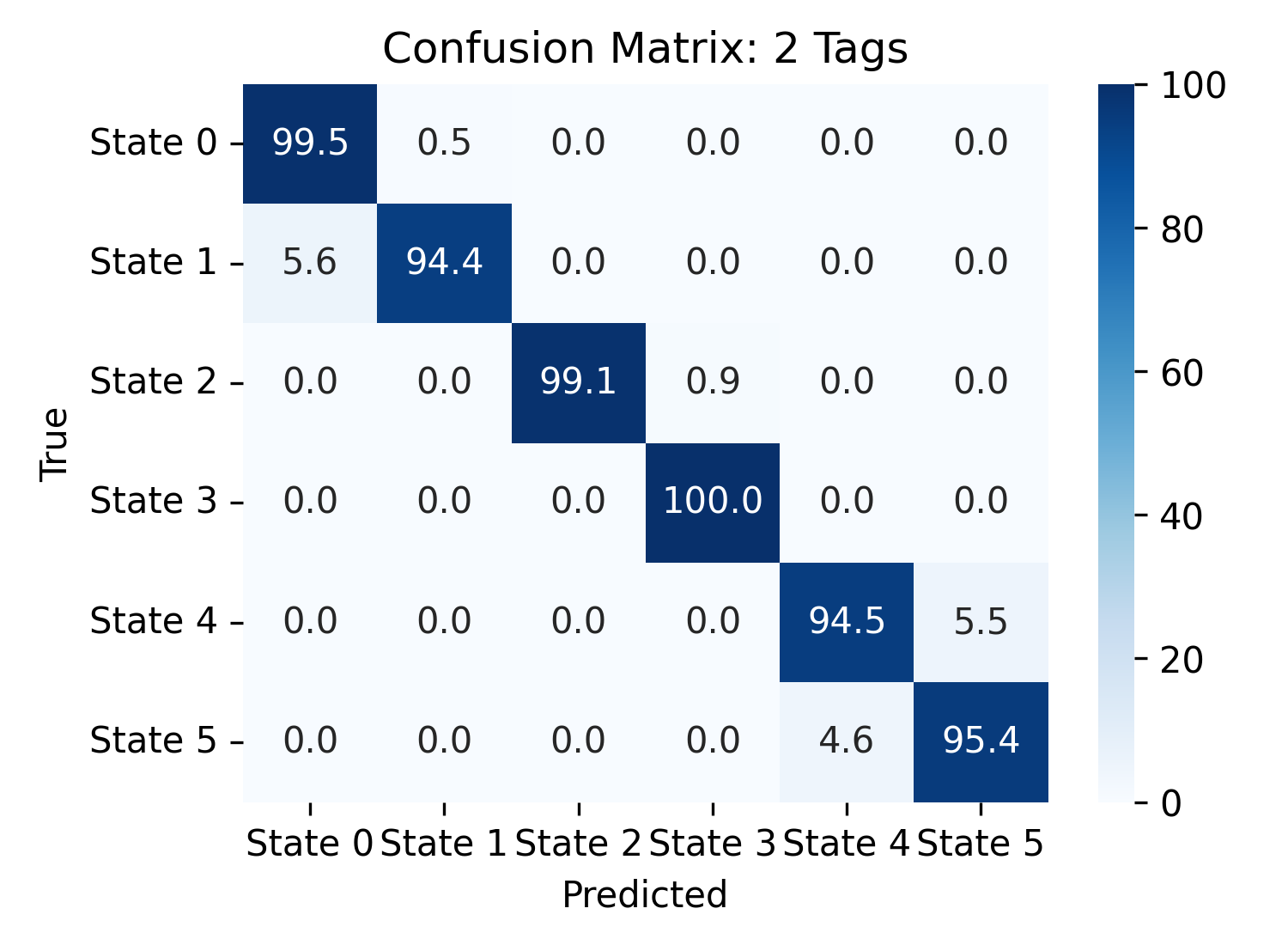}
  \caption{2 Tag Classifier Confusion Matrix}
  \Description{2 Tag Classifier Confusion Matrix}
  \label{fig:cf_side}
\end{figure}

The 2 tag scenario also performs very well, with a test accuracy of 97.13\% across all classes (Figure~\ref{fig:cf_side}). Despite the lack of a 3rd tag, state 4 and 5, which rely most on the 3rd tag for differentiation, have an accuracy of around 95\%. This indicates that even without the 3D PDoA data available, other phenomena such as the tag radiation pattern were able to provide enough information for the classifier to successfully predict the orientation. An interesting thing to note is that there is some confusion between state 0 and state 1, despite the PDoA between the tags in each state theoretically being different. This may mean that the 3rd tag's radiation pattern somewhat helped differentiate between these states, as in state 0, the tag points towards the reader, while in state 1, it points perpendicular to the reader.

\begin{figure}[htbp]
  \centering
  \includegraphics[width=\linewidth]{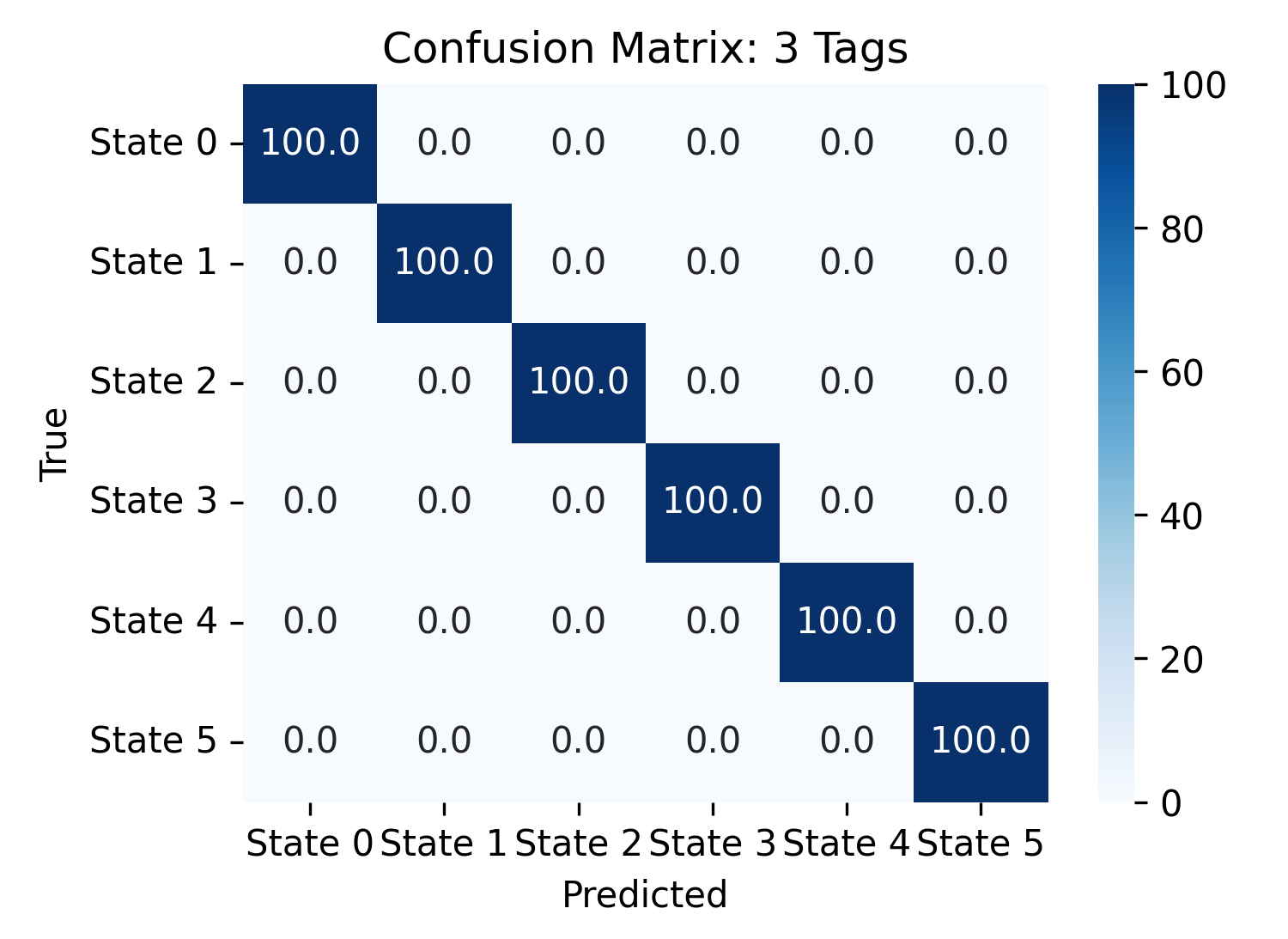}
  \caption{3 Tag Classifier Confusion Matrix}
  \Description{3 Tag Classifier Confusion Matrix}
  \label{fig:cf_rear}
\end{figure}

\subsection{Material Classification}
\subsubsection{Model Tuning}
\begin{figure}[htbp]
    \centering
    \begin{subfigure}[t]{7.5cm}
        \centering
        \includegraphics[width=\linewidth]{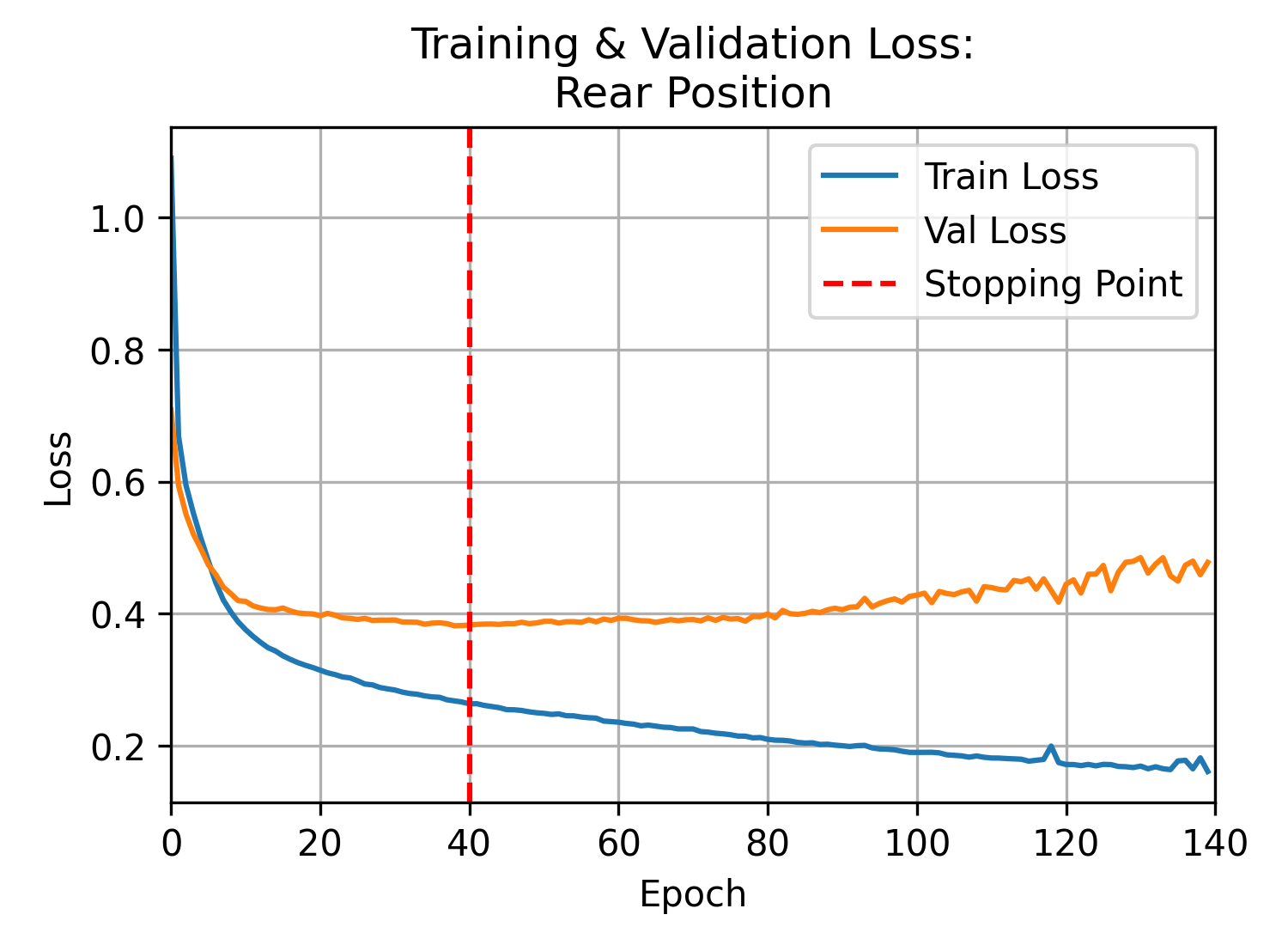}
        \caption{Rear Classifier Training \& Validation Loss}
        \Description{Rear Classifier Training \& Validation Loss}
    \end{subfigure}
    \hfill
    \begin{subfigure}[t]{7.5cm}
        \includegraphics[width=\linewidth]{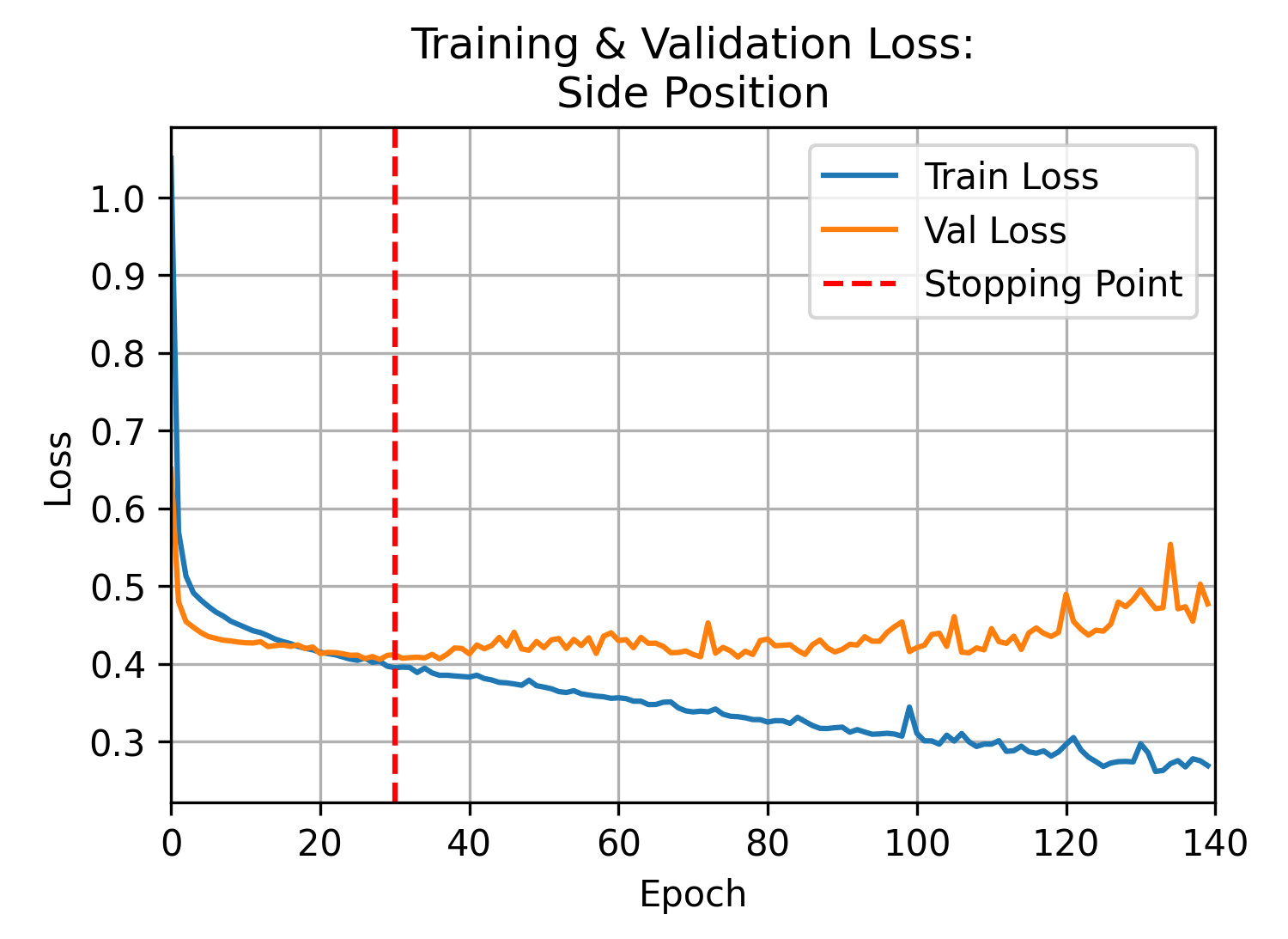}
        \caption{Side Classifier Training \& Validation Loss}
        \Description{Side Classifier Training \& Validation Loss}
    \end{subfigure}
    \caption{Neural Network Training \& Validation Loss}
    \label{fig:tv_nn}
\end{figure}

We again manually tuned hyperparameters for the feedforward neural network material classifiers, with the best settings shown in Table~\ref{tab:rf_arch}  The performance metrics achieved by both classifiers are shown in Table~\ref{tab:tv_nn}. As seen in the raw data box plots, the side tag position has much more variance in RSSI in each class from multipathing, and similar phase angle variance between classes. We found that the neural network for the side position required more hidden layers to achieve its best classification accuracy when compared to the rear tag position, likely due to the harder to differentiate data and skewed distributions. We utilized early stopping to prevent overfitting in the neural networks for each scenario. Figure~\ref{fig:tv_nn} contains the training/validation loss plots, with the stopping epoch marked.

\begin{table}[htbp]
    \centering
    \begin{tabular}{p{2.5cm}|p{2.5cm}p{2.5cm}}
        \toprule
        \textbf{ } & \textbf{Rear Classifier} & \textbf{Side Classifier} \\
        \midrule
        Features & 4 & 4 \\
        Classes & 5 & 5 \\
        Batch Size & 16 & 16 \\
        Learning Rate & 0.001 & 0.001\\
        Epochs & 40 & 30 \\
        Parameters & 12,261 & 16,421\\
        Neurons Per \\ Hidden Layer & 128, 64, 48 & 128, 64, 64, 48 \\
        \bottomrule
    \end{tabular}
    \caption{Neural Network Architecture}
    \Description{Neural Network Architecture}
    \label{tab:nn_arch}
\end{table}

\begin{table}[htbp]
    \centering
    \begin{tabular}{p{3.5cm}|p{2cm}p{2cm}}
        \toprule
        \textbf{ } & \textbf{\shortstack{Rear\\Classifier}} & \textbf{\shortstack{Side\\Classifier}} \\
        \midrule
        Stopping Epoch & 40 & 30 \\
        Training Accuracy & 0.8858 & 0.8019 \\
        Validation Accuracy & 0.8396 & 0.7985 \\
        Test Accuracy & 0.8587 & 0.7875 \\
        \bottomrule
    \end{tabular}
    \caption{Material Classifier Performance}
    \Description{Material Classifier Performance}
    \label{tab:tv_nn}
\end{table}

\subsubsection{Rear Tag Material Classification}
The rear material classifier achieves an overall test accuracy of 85.87\%, showing strong performance across all material classes. Interestingly, the chips had 100\% classification accuracy, almost certainly due to the huge difference in RSSI and phase variance caused by the metallic packaging's reflective nature. The clothes and control suffered slight confusion, likely a result of their similar median RSSI's, as the clothes have low density and high air content, resulting in less absorption. As the rear position gives the greatest tag occlusion and signal penetration, these results show that RFID based material sensing works well in this scenario.

\subsubsection{Side Tag Material Classification}
Compared to the rear material classifier, the side material classifier has a lower test accuracy of 78.75\%, with decreased performance for the toilet paper, chips, and plastic wrap. Notably, the toilet paper and plastic wrap have significant confusion and below average classification accuracy. As discussed in our raw data analysis, these two materials have a similar RSSI and phase distribution, likely due to the lower signal penetration in the side position resulting in less accumulation of differences caused by material interactions. Nearly all class pairs suffer a small amount of confusion in the side tag position, likely from multipath signals that did not penetrate through the package having similar RSSI and phase measurements. 

\begin{figure}[htbp]
  \centering
  \includegraphics[width=\linewidth]{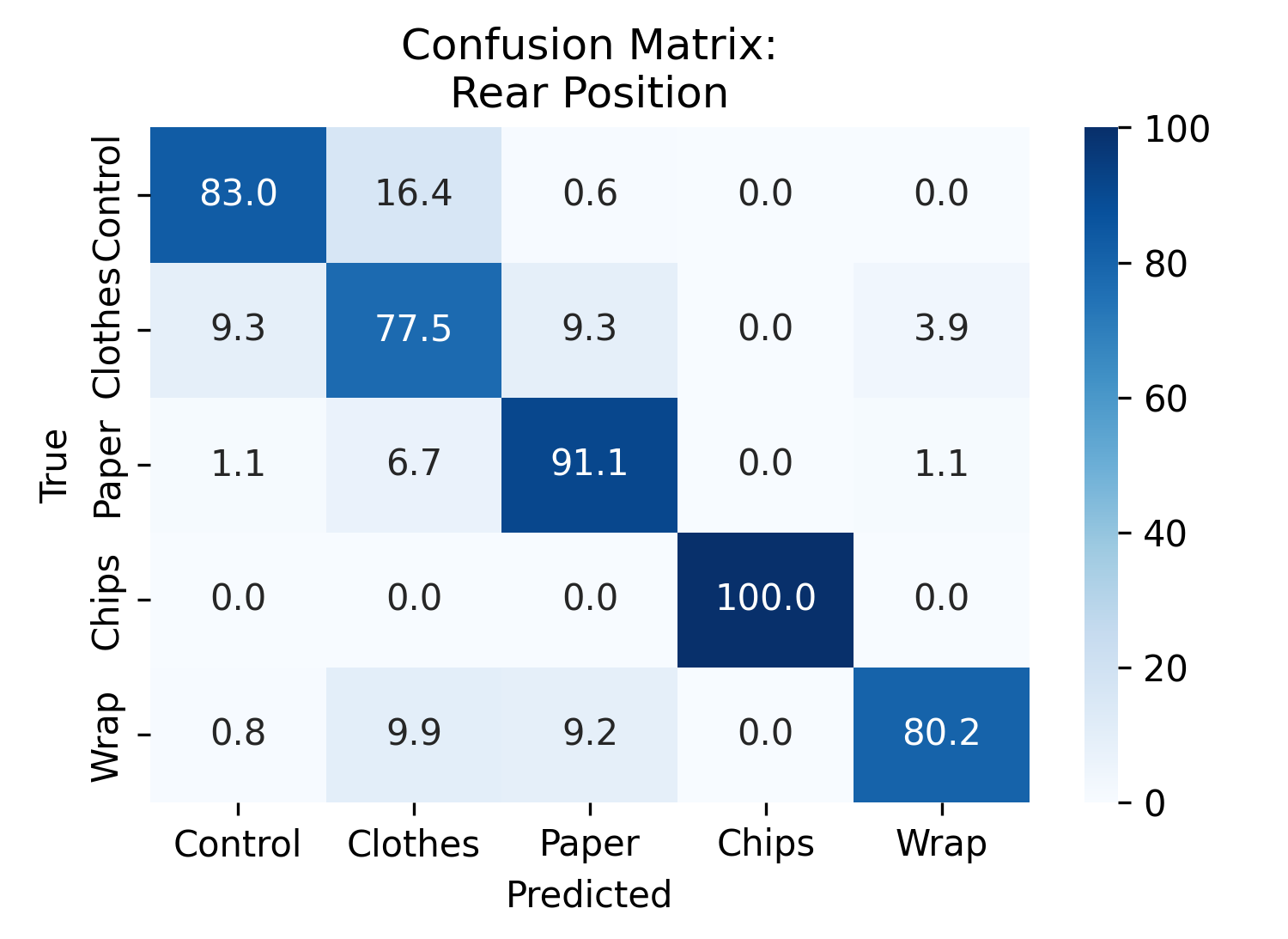}
  \caption{Rear Classifier Confusion Matrix}
  \Description{Rear Classifier Confusion Matrix}
  \label{fig:cf_rear}
\end{figure}

\begin{figure}[htbp]
  \centering
  \includegraphics[width=\linewidth]{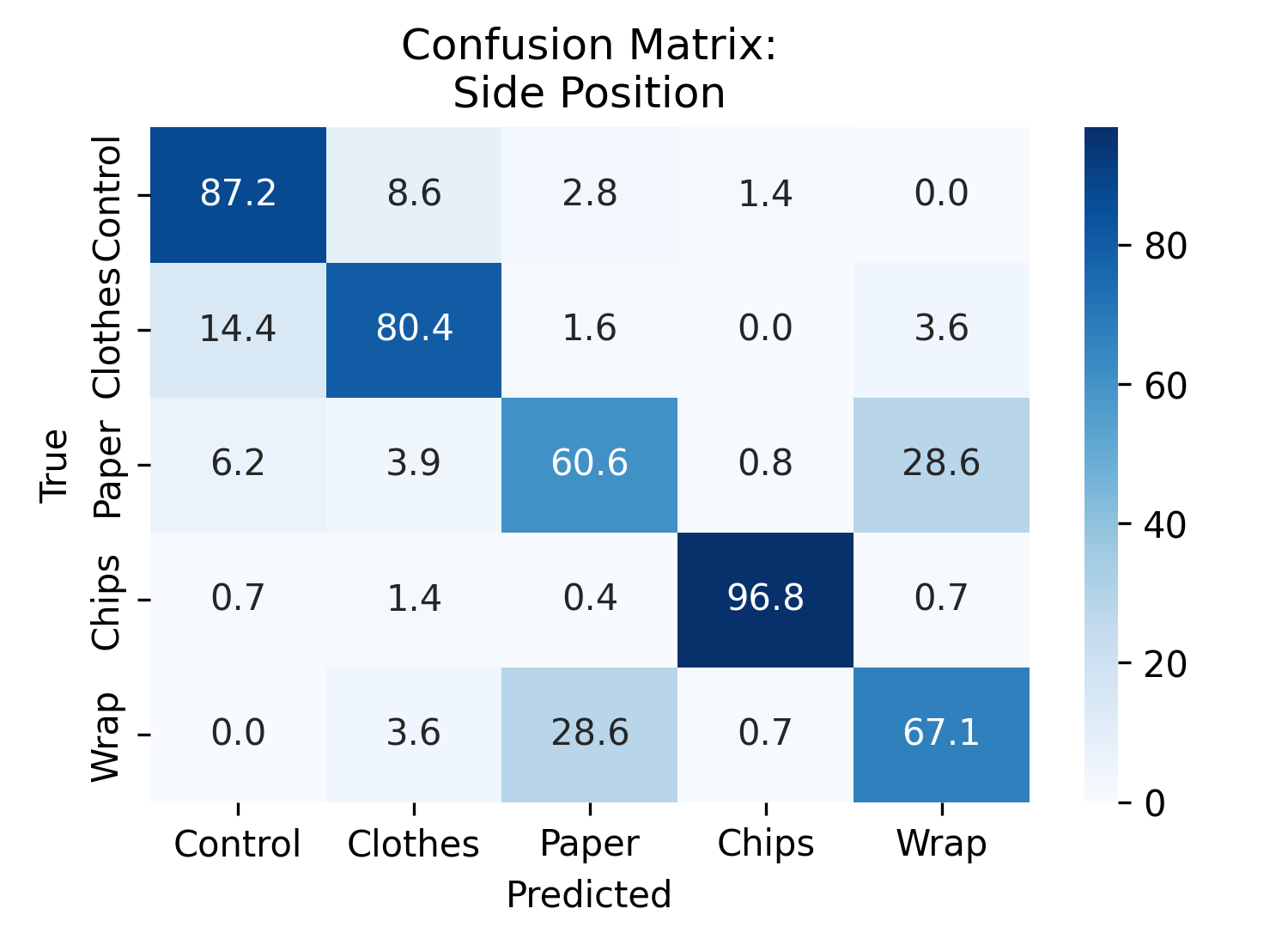}
  \caption{Side Classifier Confusion Matrix}
  \Description{Side Classifier Confusion Matrix}
  \label{fig:cd_side}
\end{figure}

\subsection{Unified Sensing Pipeline}
The full TagLabel system, shown in Figure~\ref{fig:arch}, uses the four classifiers for orientation and material sensing to create a single unified system. This system takes in the mean and variance of RSSI and phase angle for a sample with data from 2 or 3 tags, determines the orientation, and selects the appropriate material classifier based on tag positions, outputting a material prediction for the package contents. We saved the weights of each classifier after training and loaded them in the unified pipeline program to assess the performance of the full system, using a separate 1000 sample test split that was not used in the training, validation, or testing of the other classifiers. Note that this results in some differences in the final material sensing performance compared to the material classifier confusion matrices, due to this test set differing from those used to test the classifiers.

\subsubsection{Three Tag Unified Pipeline}
\begin{figure}[htbp]
  \centering
  \includegraphics[width=\linewidth]{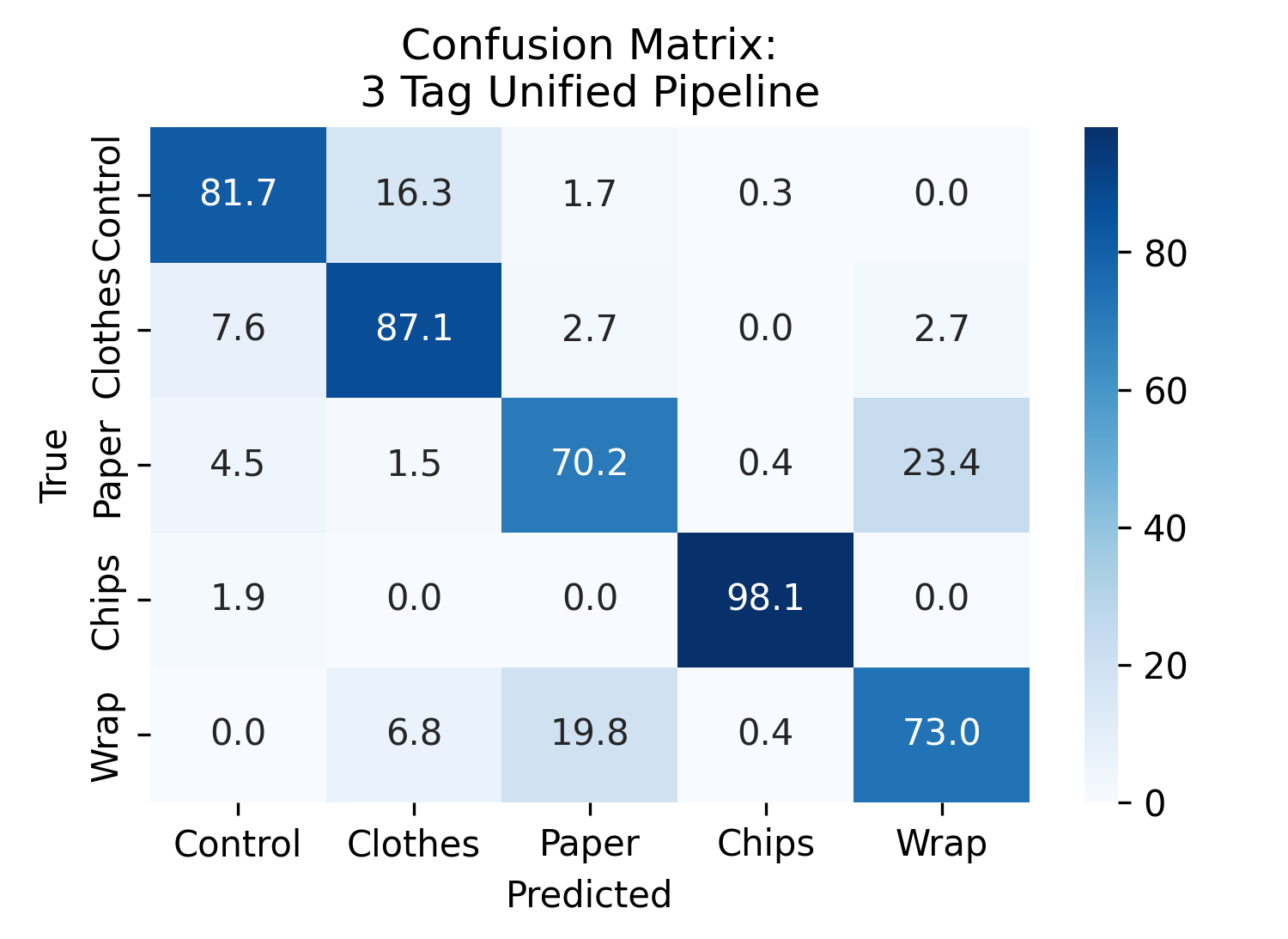}
  \caption{3 Tag Unified Pipeline Confusion Matrix}
  \Description{3 Tag Unified Pipeline Confusion Matrix}
  \label{fig:ucf_3}
\end{figure}

The 3 tag system performed strongly, with a test accuracy of 81.91\%, between that of the rear and side material classifiers. Since the 3 tag orientation classifier identifies all orientations with 100\% accuracy, these results are dependent entirely on the material classifier performance. In the 3 tag scenario, 3/6 positions use the rear classifier, while the remaining 3/6 use the side classifier, leading the results to be a mix with characteristics of both classifiers. As seen in the confusion matrix (Figure~\ref{fig:ucf_3}), some confusion still occurs between the toilet paper and plastic wrap, but to less of an extent than the side classifier's test results. Other characteristics are still present, such as the low density clothes having slight confusion with the control. 

\begin{figure}[htbp]
  \centering
  \includegraphics[width=\linewidth]{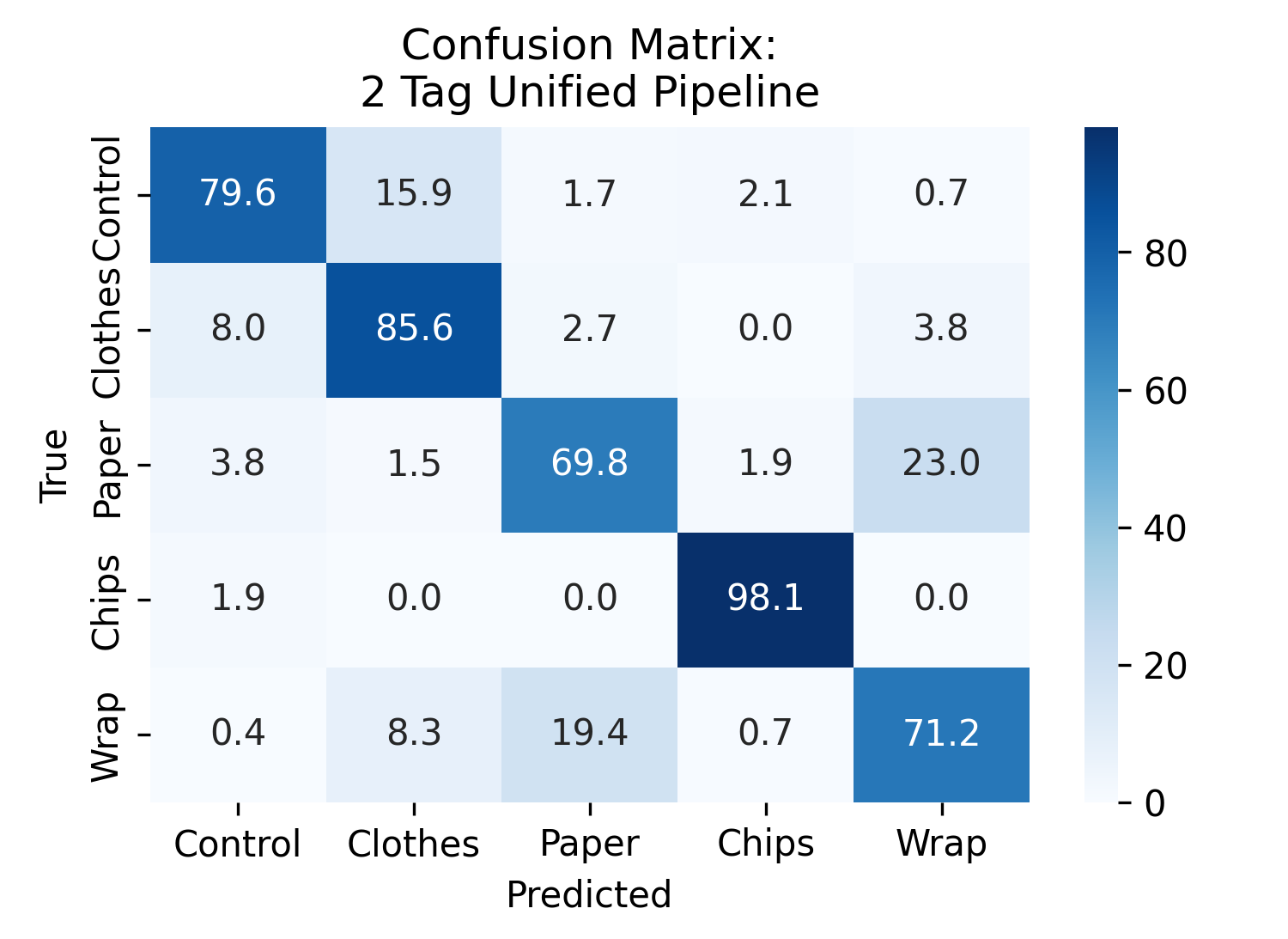}
  \caption{2 Tag Unified Pipeline Confusion Matrix}
  \Description{2 Tag Unified Pipeline Confusion Matrix}
  \label{fig:ucf_2}
\end{figure}
\subsubsection{Two Tag Unified Pipeline}
The 2 tag system also performs impressively well, with a test accuracy of 80.74\%. With the 2 tag orientation classifier, orientation sensing accuracy drops to 97\%, which likely accounts for a small reduction in performance of the unified sensing pipeline. For material sensing, 4/6 positions now use the side classifier, and 2/6 use the more accurate rear classifier, which also contributes to the small reduction in performance. Across all classes, performance remains within 1-2\% of the 3 tag pipeline, indicating that a 2 tag solution can perform nearly equal to that of the 3 tag system, while being easier to implement.

\section{Discussion}
\subsection{Two vs. Three Tags}
In both the orientation classification and the full TagLabel classification pipeline, the 2 tag scenario offers strong performance, within 1-3\% of the 3 tag versions. Even without 3D phase differences available, the random forest classifier was able to leverage deeper patterns to maintain high levels of accuracy, such as the radiation patterns of the tags and the bottom tag being occluded. Additionally, although the side classifier performs worse than the rear classifier and is used 2/3 times in the 2 tag scenario, the combination of both classifiers is still able to provide an 80.7\% sensing accuracy, delivering a very reasonable level of performance. Given that the 2 tag solution is easier to integrate into existing shipping systems, we would recommend using 2 tags going forward.
\subsection{Package to Reader Distance}
The center of the package was placed 50 cm from the RFID reader in all of our tests. However, as RSSI drops off with distance, it may be more optimal to place the package closer to the reader. This also increases the angle between tags in the side position and the reader, which could potentially improve material sensing performance. At the same time, stronger RFID signals may be attenuated less, causing material sensing accuracy to decrease with thinner materials. Material sensing accuracy at varying distances is something that needs to be studied further for any possible commercial deployments of this technology.
\subsection{Tag Placement}
For our experiments, the tags were placed on the center of each face of the package, with the tags attached in the orientations shown in Figure~\ref{fig:orient}. As seen in our evaluation, when the tag is in the side position, it has lower signal penetration through the package compared to the rear position, resulting in decreased accuracy. It may be advantageous to place the tags at the edges of the packages or on different faces than the ones we used in order to increase signal penetration in the side positions. Additionally, as we saw with the 2 vs 3 tag scenarios, the antenna radiation patterns are able to make up for the loss of a third tag. By placing the tags in varying rotations on the package such that their antenna pattern is aligned with the reader, it may be possible to achieve improved material sensing performance from all positions. 

\section{Conclusion}
This study demonstrates that RFID based sensing can provide reliable insight into both the material and orientation of packages using only low cost passive UHF tags. By measuring how different materials alter RFID signals, we show that RSSI and phase angle provide enough information to support accurate material classification when paired with machine learning. At the same time, orientation can be inferred from PDoAs, antenna gain patterns, and tag occlusion effects, enabling the system to identify the orientation of packages without the need for complex analytic models. TagLabel combines both orientation and material sensing to solve a major challenge in RFID based inspection: the same tag can behave differently depending on where it is located or how the package is rotated. By leveraging orientation to pick the best tag for material sensing, the system maintains consistent accuracy, up to 82\%, with either two or three tags.

Integrating RFID based package content screening into logistics workflows can expand the role of RFID beyond just identification. Systems scanning boxes with smart RFID shipping labels can check package contents without opening them, flag suspicious or prohibited items, and automate orientation detection tasks, improving safety and reducing labor requirements. As the sensing process operates in seconds and uses off the shelf UHF tags common in retail supply chains, it can easily be integrated into existing conveyor and sorting infrastructure. Combining material and orientation data makes this approach practical for real world deployments and offers a path toward faster, safer, and more automated logistics operations.


\bibliographystyle{ACM-Reference-Format}
\bibliography{references}

\end{document}